\documentclass[pra,superscriptaddress,showpacs,preprint,aps,longbibliography]{revtex4-1}
\usepackage{epsfig}             
\usepackage{epstopdf}           
\usepackage{hyperref}           
\usepackage{color}
\usepackage{colortbl}
\usepackage{graphicx}
\usepackage{amssymb,amsmath}
\usepackage{subfigure}
\usepackage{caption}
\usepackage{enumerate}
\usepackage{gensymb}
\usepackage{url}

\begin{document}

\title[ATI]{Above-Threshold Ionization and Laser-Induced Electron Diffraction in Diatomic Molecules}

\author{Noslen Su\'arez}
\email[]{noslen.suarez@icfo.es}
\affiliation{ICFO - Institut de Ci\`encies Fot\`oniques, The Barcelona Institute of Science and Technology, Av. Carl Friedrich Gauss 3, 08860 Castelldefels (Barcelona), Spain}

\author{Alexis Chac\'on}
\affiliation{ICFO - Institut de Ci\`encies Fot\`oniques, The Barcelona Institute of Science and Technology, Av. Carl Friedrich Gauss 3, 08860 Castelldefels (Barcelona), Spain}

\author{Marcelo F. Ciappina}
\affiliation{Max-Planck-Institut f\"ur Quantenoptik, Hans-Kopfermann-Str. 1, 85748 Garching, Germany}
\affiliation{Institute of Physics of the ASCR, ELI-Beamlines, Na Slovance 2, 182 21 Prague, Czech Republic}

\author{Benjamin Wolter}
\affiliation{ICFO - Institut de Ci\`encies Fot\`oniques, The Barcelona Institute of Science and Technology, Av. Carl Friedrich Gauss 3, 08860 Castelldefels (Barcelona), Spain}

\author{Jens Biegert}
\affiliation{ICFO - Institut de Ci\`encies Fot\`oniques, The Barcelona Institute of Science and Technology, Av. Carl Friedrich Gauss 3, 08860 Castelldefels (Barcelona), Spain}
\affiliation{ICREA - Instituci\'{o} Catalana de Recerca i Estudis Avan\c{c}ats, Lluis Companys 23, 08010 Barcelona, Spain}

\author{Maciej Lewenstein}
\affiliation{ICFO - Institut de Ci\`encies Fot\`oniques, The Barcelona Institute of Science and Technology, Av. Carl Friedrich Gauss 3, 08860 Castelldefels (Barcelona), Spain}
\affiliation{ICREA - Instituci\'{o} Catalana de Recerca i Estudis Avan\c{c}ats, Lluis Companys 23, 08010 Barcelona, Spain}

\date{\today}

\begin{abstract}
Strong field photoemission and electron recollision provide a viable route to extract electronic and nuclear dynamics from molecular targets with attosecond temporal resolution. However, since an {\em ab-initio} treatment of even the simplest diatomic systems is beyond today's capabilities approximate qualitative descriptions are warranted. In this paper, we develop such a theoretical approach to model the photoelectrons resulting from intense laser-molecule interaction. We present a general theory for symmetric diatomic molecules in the single active electron approximation that, amongst other capabilities, allows adjusting both the internuclear separation and molecular potential in a direct and simple way. More importantly we derive an analytic approximate solution of the time dependent Schr\"odinger equation (TDSE), based on a generalized strong field approximation (SFA) version. Using that approach we obtain expressions for electrons emitted transition amplitudes from two different molecular centres, and accelerated then in the strong laser field.  In addition, our approach directly underpins different underlying physical processes that correspond to i) direct tunnelling ionization; ii) electron rescattering on the centre of origin; iii) and, finally, electron rescattering on a different centre. 
One innovative aspect of our theory is the fact that the dipole matrix elements are free from non-physical gauge and coordinate system dependent terms -- this is achieved by adapting the coordinate system, in which SFA is performed, to the centre from which the corresponding part of the time dependent wave function originates.
Our analytic results agree very well with the numerical solution of the full three-dimensional TDSE for the H$_2^+$ molecule. 
Moreover, the theoretical model was applied to describe laser-induced electron diffraction (LIED) measurements of O$_2^+$ molecules, obtained at ICFO, and reproduces the main features of the experiment very well. Our approach can be extended in a natural way to more complex molecules and multi-electron systems.
\end{abstract}

\maketitle

\section{Introduction}


\subsection{Imaging in strong fields: high-order harmonic generation (HHG) and laser-induced electron diffraction (LIED)}

One of the most exciting prospects of strong field and attosecond physics is the extraction of electronic and nuclear information on the attosecond temporal and picometer spatial scales~\cite{Corkum1993}. Strong field techniques such as high harmonic spectroscopy
(HHS)~\cite{BakerScience2006, MairessePRL2008,Woerner2010} exploit the quiver motion of an electron which is liberated from the target structure itself and analyze either the recombination spectrum or the momentum distribution of the rescattering electron~\cite{Zuo1996,LeinJPB2007,Meckel2014}


In a seminal work, Villeneuve {\it et al.} \cite{NatItatani2004} demonstrated that a tomographic reconstruction from a HHS measurement yields the Dyson orbital of N$_2$~\cite{SergeiPRL2006}. The original interpretation of these experiments was based on the strong field approximation (SFA) description of the process~\cite{Lewenstein1994}, which provides a fully quantum description of the well-known ``three step model"~\cite{Corkum1993, Kulander1992a,Kulander1993, Smirnova2014}. Villeneuve et al.'s~\cite{NatItatani2004} experiment has triggered a true avalanche of experimental and theoretical works on the subject~\cite{SmirnovaNature2009, BakerScience2006, Haessler2010, McFarland2008, Diveki2012, Wang2016, MorishitaPRL2016, Meckel, Milosevic2006}. The use of approximations (and in particular the SFA) in the tomography of molecular orbitals is, however, under permanent debate and full for controversial issues: the results strongly depend on the gauge, the choice of the dipole radiation form, the molecular orbital symmetry and degree of alignment and the reconstruction axis (cf.~\cite{JensPRA2008, MilosevicPRA2009,Li2013, Qin2012, Senad2012,Nguyen201112,Jens2011, CarlaPRA2010, Jens2010}). 

LIED~\cite{Zuo1996,Lein2002,Lin2010} is based on extracting structural information directly from electrons which are elastically scattered of the parent ion. Recently, LIED has been used to successfully recover structural information from diatomic and larger polyatomic molecules~\cite{Meckel,blaga2012,Pullen2015,Ito2016}. An electron may directly depart from the molecule and contribute to the lower energy region of the above-threshold ionization (ATI) spectrum --this process is termed direct tunneling-or it might return to the target molecular ion, driven by the still present laser electric field, and rescatter, thereby gaining much more energy. This high energetic electron could excite the remaining ion or even cause the detachment of a second electron (for a consistent description of these processes within the framework of the SFA and Feynman's path-integral approach see \cite{MaciekScience2001}).



The viability of this self-imaging technique to retrieve structural information of molecular and atomic systems has been demonstrated in a series of contributions~\cite{Xu2014,blaga2012, Pullen2015}. The idea is here to gain insight about the electronic structure of molecular targets interpreting the energy spectra and angular distribution of above-threshold ionization electrons. In particular the high energy region of the ATI spectra-which is mainly due to the rescattering process-is particularly sensitive to the structure of the target. I.e. the rescattered electron has incurred information about the target it rescattered off and hence permits extracting structural information.

Some efforts have been already made in the study and development of new theoretical tools to investigate the structure of complex systems (such as molecules, atom clusters and solids) using the ATI spectra. Amongst those investigations, ATI from diatomic systems is the most widely studied process ~\cite{MilosevicPRA2006,LeinPRL2012,CarlaPRA2007,Marcelo2007}. Two methods are commonly used and compared: a fully quantum-mechanical description based on the numerical solution of the time dependent Schr\"odinger equation (TDSE) and approximated methods based on the SFA and other quasi-classical approaches. We should mention, however, that the former is only  feasible for simple diatomic molecules, e.g.~H$_2^+$, H$_2$, D$_2$,  and within the single active electron (SAE) approximation. The results so far focus, for instance, in the differences between the length and velocity gauge~\cite{LeinPRA2006}, the influence of the internuclear distance~\cite{CarlaPRA2007} or alignment~\cite{ LeinJPB2007, LeinPRA2002, LeinPRL2012, MilosevicPRA2006, MilosevicPRL2008} on the ATI photoelectron spectra and the importance of the residual Coulomb interaction~\cite{Marcelo2007,Marcelo2009}.
While TDSE provides the most accurate description of the underlying physics behind LIED, it is numerically and computationally very costly. In addition, the multidimensional TDSE can currently not be solved for complex molecules, multi-electron systems or molecular systems evolving in time. Thus, one has to resort to approximate descriptions such as the SFA and related methods to adequately describe the more complex instances of LIED. Similar arguments can be put forward for molecular orbital tomography methods based on HHG. 

\subsection{Overcoming the drawbacks of SFA}

The standard SFA method, however, has severe drawbacks, namely the electronic states in the continuum are described in their simplest approximation by Volkov states-a plane wave in the presence of the laser field-or, in a slightly more sophisticated version, by Coulomb-Volkov states or similar ones which take into account Coulomb corrections~\cite{Bauer1997,Popruzhenko, PotvliegeBook}. These states are typically not orthogonal to the target bound states and this introduces spurious contributions. For instance, when we compute the transition dipole matrix element ${\bf d}(\bf{v})$ between the bound $|{0}\rangle$ and continuum $|{\phi_{\bf v}}\rangle$ states, the results depends linearly on the choice of the center of the coordinate system: 
${\bf d}({\bf v})=q_e\langle \phi_{\bf v}| \hat {\textbf {r}}| {0}\rangle \neq q_e\langle \phi_{\bf v}|(\hat{\textbf{r}} -\textbf{R}) | 0 \rangle,$
where ${\textbf R}$ is a constant coordinate shift, typically corresponding to the distance between the nuclei-the so-called internuclear distance-in a two-center molecule. This is an artificial and nonphysical effect, particularly problematic when  ${\textbf R}\to \infty$. Most authors handle this problem by neglecting the linear terms in ${\textbf R}$ in the dipole matrix elements~\cite{CarlaPRA2007,LeinPRA2006,MilosevicPRL2008}. Nevertheless, this is not a systematic approach, since it does not solve adequately problems related with various phase factors appearing on the molecular dipole matrix elements.  In addition, they present a strong dependence on the choice of the gauge, or problems with the correct asymptotic behavior for ${\textbf R}\to \infty$ and yet to ${\textbf R}\to 0$ (cf.~\cite{CarlaPRA2007}). Furthermore, the agreement with the TDSE results is typically poor.  Besides of the mentioned weaknesses, we should note that these previous studies have led to a relatively good description of the ATI process in diatomic molecules.

In this paper we propose a natural and systematic solution of all the above mentioned problems by extending the SFA to complex molecules without ambiguities. Our version of the SFA for ATI and HHG has the following appealing properties: 

{\Roman{enumii}}
\begin{enumerate}[(a)]
\item It analytically reproduces the results for ${\textbf R}\to \infty$; for the particular case of diatomic molecules this corresponds to two identical atoms (sources) generating electronic (photonic) states with a phase difference corresponding to the distance ${\textbf R}$ between them. 

\item It reproduces analytically the asymptotic limit for ${\textbf R}\to 0$; for the case of a diatomic molecule we end up with the usual single atom formulation. 

\item Statements (a) and (b) agree well with their counterpart solutions obtained using the 3D-TDSE.

\item It allows us to interpret the results in terms of quantum orbits, e.g.~we could disentangle contributions for electrons originating at a given center $\textbf {R}_i$ that rescatters at another one $\textbf {R}_j$, etc.. 

\item It is free of nonphysical dependencies on $\textbf{R}_i$.  

\item It agrees well with experimental results at ICFO concerning O$_2^+$ molecules.  
\end{enumerate}
Our approach is based on the following observation: both in the ATI and HHG cases the molecular response, which is determined by the probability amplitude of an electron in the continuum with a given energy and velocity, depends linearly on the wavefunctions of the initial (ground) state. More generally, the solution of the {\it linear} TDSE depends {\it linearly} on the wavefunction of the initial state. Commonly, for a molecule it is natural to write this function a summation of the contributions corresponding to different nuclei -- for a diatomic molecule it is a sum of two terms, for triatomic molecules a sum of three, etc. Following this reasoning, our modified SFA consists in the following steps:

\begin{itemize}
\item Decompose the initial ground  state of the molecule into a superposition of terms centered at $\textbf {R}_i$, $i=1,2,\dots$, i.e.~at the position where the heavy nuclei are located.

\item Solve independently a TDSE, exactly or using the SFA for each term, using a coordinate system centered at each $\textbf{R}_i$. 

\item Transform at the end all terms to the same coordinate system and coherently add them up. 

\end{itemize}
Further, this approach is formally exact, as the exact numerical solutions of the TDSE are used; in fact it might even has some numerical advantages. On the other hand, the formulation is approximated if the SFA  is used to solve the TDSE -- but this approach seems to give particularly robust outcomes which agreed very well with the exact ones. 

We illustrate our point with the simplest possible example: a two-center molecule with two identical atoms separated at certain distance $R$ driven by a strong ultrashort laser field linearly polarized along the $z$-direction, but nothing prevents to apply our formalism to more complex molecular targets. To model the electron-heavy ion interactions we take advantage of the short-range potential model developed by Becker et al.~\cite{Becker1990}.  Considering the ATI spectra is sensitive to the internuclear distance and the orientation between the molecular axis and the polarization direction of the laser field $z$, we aim to generalize the SFA for atomic systems presented in~\cite{Lewenstein1995, PRANoslen2015} to the above mentioned diatomic molecular targets. Furthermore, we put particular emphasis to all the possible scenarios: tunneling ionization from both centers and propagation in the continuum till the measurement process; tunneling ionization from one particular center, electron propagation in the continuum and rescattering on the same parent center; tunneling ionization from one center, electron propagation in the continuum and rescattering with its neighboring parent center; and (perhaps the most peculiar one) tunneling ionization from one center, electron propagation in the continuum and rescattering on the same center, causing electron rescattering from the parent neighboring center.

We stress out, in the end, that the agreement of our results with the TDSE solution is remarkably good, which allows us to seriously think in extensions of our SFA version  to tackle more complex molecules, with 3 or more molecular centers, and systems with more active electrons, where the solution of the exact Schr\"odinger equation is not currently available. 
\vspace{-.2cm}
\subsection{Plan of the paper}

This article is organized as follows. In Sec.~II, we write the formulae for the two-center molecular system ATI transition amplitude, for both the direct and rescattered electrons, using the prescriptions presented above. In Sec.~III we introduce a particular nonlocal short-range (SR) model potential to calculate the bound and rescattering electron states. The matrix elements that describe the ionization and rescattering processes are then provided in an analytic form. We use them in Sec.~IV to compute both energy-resolved ATI and two-dimensional electron momentum distributions for a diatomic molecule. Here our numerical results are compared with numerical  results obtained from TDSE calculations. The basic analysis of the interference minima of the photoelectron spectra and the discussion of how structural information, the internuclear distance, could be retrieved is presented in this section.  In Sec.~V we confront and compare our results with experimental results obtained at ICFO for O$_2^+$ molecules. Clearly, the comparison is very encouraging an suggest that our theory is on the right track to describe experiments in more complex systems. 
Finally, in Sec.~VI, we summarize the main ideas and present our conclusions.

\section{Generalized Strong field approximation: transition probability amplitudes}\label{cap:Background}

\subsection{Basics of SFA: a remainder}

We aim to extend the SFA from atomic systems presented Ref.~\cite{PRANoslen2015} to molecular targets. In particular, we focus ourselves on calculating the final photoelectron spectrum by means of solving the TDSE for a molecule with two identical centers separated by a distance ${R}$ and driven by a short and intense linearly polarized laser pulse. We define the relative vector position ${\bf R}=\textbf{R}_{2}-\textbf{R}_{1}$ and one atom is placed on the $\textit{Left}$ at $\textbf{R}_{1} = -\frac{\textbf{R}}{2}$ meanwhile the other one is located on the $\textit{Right}$ at $\textbf{R}_2 = +\frac{{\bf R}}{2}$. In general, as the molecular nuclei are much heavier than the electrons and the laser pulse duration is shorter than the nuclei vibration and rotational dynamics, we fix the nuclei positions and neglect the repulsive interaction between them. Further,  throughout the formulation we consider the so-called Single Active Electron (SAE) approximation.
 
The TDSE that describes the whole laser-molecule interactions (atomic units are used throughout this paper unless otherwise stated) can be written as:
\begin{eqnarray}
\nonumber
i\frac{ \partial}{ \partial t} | \Psi(t) \rangle&=&\hat{H} | \Psi(t) \rangle, \\ 
&=&[\hat{H}_0 + \hat{V}_{int}({\bf r},t)]| \Psi(t) \rangle,
\label{Eq:SE}
\end{eqnarray}
where $\hat{H}_0=\frac{{\hat{\bf p}}^2}{2} + \hat{V}(\textbf{r})$ defines the laser-field free hamiltonian,  with ${\hat{\bf p}}=-i{\nabla}$ the canonical momentum operator and $\hat{V}(\textbf{r})$ potential operator that describes the interaction of the nuclei with the active electron and $\hat{V}_{int}({\bf r},t)=-q_e\hat{{\bf E}}(t)\cdot\hat{{\bf r}}$ represents the interaction of the molecular system with the laser radiation, written in the dipole approximation and length gauge. $q_e$ denotes the electron charge which in atomic units has the value of $q_e=-1.0$~a.u. Finally, the linearly polarized, in the $z$-axis, laser electric field has the form:
 $\textbf{E}(t) = \mathcal{E}_0\:f(t) \sin(\omega_0 \:t + \phi_0)\,{\bf e}_{z}$, where $\mathcal{E}_0, \omega_0, f(t)$ and $\phi_0$ are the electric field peak amplitude, the carrier frequency, the laser envelope and the carrier-envelope phase (CEP), respectively. We have defined the laser pulse envelope as $f(t)=\sin^2(\frac{\omega_0t}{2N_c})$ where $N_c$ is the number of total cycles.
 
We shall restrict  our model to  the  low ionization  regime, where the  SFA is valid~\cite{ Keldysh1965, Faisal1973,Reiss1980,Lewenstein1986, Lewenstein1994,Lewenstein1995}.  Therefore, we work in the tunneling regime, where the Keldysh parameter  $\gamma=\sqrt{I_p/2U_p}$ ($I_p$ is the ionization potential of the system and $U_p=\frac{\mathcal{E}_0^2}{4\omega_0^2}$ the ponderomotive energy acquired by the electron during its incursion in the field) is less than one, i.e.~$\gamma<1$. In addition,  we assume that $V(\textbf{r})$  does not  play an important role in the electron  dynamics  once the electron  appears in the continuum. 

These observations,  and the following three statements, define the standard SFA, namely:
\begin{enumerate}[(i)]
\item Only the ground state, $|0 \rangle$, and the continuum  states, $ |\textbf{v}\rangle$, are taken into account in the interaction process.
\item There is no depletion of the ground state $(U_p < U_{sat})$.
\item The continuum  states  are approximated by Volkov states; in the continuum the electron is considered as a free particle solely moving in the laser electric field. 
\end{enumerate}
For a more detailed discussion of the validity of the above statements see e.g.~Refs.~\cite{Lewenstein1994,Lewenstein1995, PRANoslen2015}. 

\subsection{SFA: an appropriate treatment of two centre systems}

Based on the statement (i), and the linearity of the Schr\"odinger equation we propose a general state for the system: 
\begin{eqnarray}
|\Psi(t) \rangle =|\Psi_{\mathcal L}(t) \rangle + |\Psi_{\mathcal R}(t) \rangle, \label{StateEq}
\end{eqnarray}
which is the coherent superposition of two states $|\Psi_{\mathcal L}(t) \rangle$ and  $|\Psi_{\mathcal R}(t) \rangle$. The sub-indexes `${\mathcal L}$' and `${\mathcal R}$' refer to the contributions of the spatially localized \textit{Left} and \textit{Right} nuclei, respectively. We note that those \textit{Left-Right} states are not orthogonal between them. \\
Following the same assumption that in our previous contribution~\cite{Lewenstein1995, PRANoslen2015}, each single state can be written as the coherent superposition of ground \textit{Left-Right} and continuum states
\begin{equation}
| \Psi_{\mathcal L} (t)\rangle= e^{\textit{i}I_p\textit{t}}\bigg(a(t) |0_{\mathcal L} \rangle + \: \int{\textit{d}^3 \textbf{v} \:  \textit{b}_{\mathcal L}( \textbf{v},t) |\textbf{v}\rangle} \bigg),
\label{Eq:PWavefL}
\end{equation}
\begin{equation}
| \Psi_{\mathcal R}(t) \rangle= e^{\textit{i}I_p\textit{t}}\bigg(a(t) |0_{\mathcal R} \rangle + \: \int{\textit{d}^3 \textbf{v} \:  \textit{b}_{\mathcal R}( \textbf{v},t) |\textbf{v}\rangle} \bigg).
\label{Eq:PWavefR}
\end{equation}
Note that the whole ground state, i.e.~$|0\rangle=|0_{\mathcal L}\rangle + |0_{\mathcal R}\rangle$, is a composition of \textit{Left} $|0_{\mathcal L}\rangle$ and \textit{Right} $|0_{\mathcal R} \rangle$ contributions. In this way we are able to separate the whole state $| \Psi (t)\rangle$ both as the \textit{Left-Right} states described in Eq.~(\ref{StateEq}) and the two above ones (Eqs.~(\ref{Eq:PWavefL}) and (\ref{Eq:PWavefR})).

The factor, $a(t)$, represents  the  amplitude of the state $|0\rangle$ and it is considered constant in time, $a(t) \approx 1$,  under the assumptions of the statement  (ii). The pre-factor $e^{\textit{i}I_p\textit{t}}$ describes the accumulated  electron energy in the ground state where $I_p=-E_0$ ($E_0$ is the molecular ground-state energy). Furthermore, the continuum states transition amplitudes $\textit{b}_{\mathcal L}(\textbf{v},t)$ and $\textit{b}_{\mathcal R}(\textbf{v},t)$ are referring to the electron-wavefunction ionized from the \textit{Left} and \textit{Right} nuclei, respectively. 

Our main task will be thereby to derive general expressions, by means of the Eq.~(\ref{Eq:SE}) and the definitions of Eqs.~(\ref{Eq:PWavefL}) and (\ref{Eq:PWavefR}), for the transition amplitudes ${b}_{\mathcal L}({\bf v},t)$ and ${b}_{\mathcal R}({\bf v},t)$. We shall consider that $\hat{H_0}|0_{\mathcal L,R}\rangle= -I_p |0_{\mathcal L,R}\rangle$ and $\hat{H_0}
|\textbf{v}\rangle = \frac{\textbf{v}^2}{2}|\textbf{v}\rangle$ fulfil for the bound and continuum states, respectively. Consequently, the evolution of the transition amplitude ${b}_{\mathcal L}({\bf v},t)$ becomes:
\begin{eqnarray}
i\int{\textit{d}^3  \textbf{v} \:\dot{b}_{\mathcal L}( \textbf{v},t)\: |\textbf{v} \rangle} &=& \,\, \int{\textit{d}^3 \textbf{v} \bigg(\frac{\textbf{v}^2}{2}+ I_p \bigg)\textit{b}_{\mathcal L}( \textbf{v},t) |\textbf{v}}\rangle+\textbf{E}(t) \cdot  \textbf{r} |0_{\mathcal L} \rangle \nonumber \\
 && +\: \textbf{E}(t) \cdot \textbf{r}\int{\textit{d}^3 \textbf{v} \textit{b}_{\mathcal L}( \textbf{v},t)|\textbf{v} \rangle}.
  \label{Eq:TempEq}
\end{eqnarray}
On the above equation we have assumed that  the electron-nuclei  interactions are neglected once the electron appears in the continuum, based on the statement (iii). Therefore, by multiplying Eq.~(\ref{Eq:TempEq}) by $\langle {\bf v}'|$ and after some algebra, the time variation of the transition amplitude reads:
\begin{eqnarray}
\dot{b}_{\mathcal L}( \textbf{v},t) &=& \,\,  -i\bigg(\frac{\textbf{v}^2}{2}+ I_p \bigg) \textit{b}_{\mathcal L}( \textbf{v},t) + i \: \textbf{E}(t) \cdot\textbf{d}_{\mathcal L}( \textbf{v})\nonumber \\
&& -i \: \textbf{E}(t) \cdot \int{\textit{d}^3 \textbf{v}^{\prime}\textit{b}_{\mathcal L}( \textbf{v}^{\prime},t)\langle \textbf{v} | \textbf{r}  |\textbf{v}^{\prime}\rangle}.
\label{Eq:Newbl}
\end{eqnarray}
The first term on the right-hand of Eq.~(\ref{Eq:Newbl}) represents the  phase evolution  of the  electron within the oscillating laser field. In the second term, we have defined the bound-free transition dipole matrix element as:
\begin{equation}
\textbf{d}_{\mathcal L}( \textbf{v}) = - \langle \textbf{v} |\textbf{r}|0_{\mathcal L} \rangle. 
 \label{Eq:DM1}
\end{equation}
Here, $ |\textbf{v} \rangle$ represents in general an scattering state built up as the superposition of a plane wave, $| {\bf v}_p \rangle$ and corrections on the \textit{Left}, $| \delta{\bf v}_{\mathcal L} \rangle$, and on the \textit{Right}, $|\delta{\bf v}_{\mathcal R} \rangle$
\begin{equation}
 |\textbf{v} \rangle=| {\bf v}_p \rangle+| \delta{\bf v}_{\mathcal L} \rangle+| \delta{\bf v}_{\mathcal R} \rangle.
 \label{Eq:SCS}
\end{equation}
 Based on statement (iii) our formulation only considers the continuum state as a plane wave $| {\bf v}_p \rangle$ for the calculation of the bound-free dipole matrix element. 
We shall pay special attention to the computation of Eq.~(\ref{Eq:DM1}). Let us stress the fact that plane waves are not orthogonal to the bound states. Notice also that our bound state is defined depending on the relative position of one the atoms, $\textbf{R}_1$ with respect to the origin of coordinates. In this sense we need to introduce a `position correction' on the dipole transition matrix in order to avoid nonphysical linear terms dependence on $\bf{R}$ (see Section I.C for more details). So, for the \textit{Left} contribution we introduce a correction to the dipole matrix element as
\begin{eqnarray}
\textbf{d}_{\mathcal{ L}}( \textbf{v}) &=& - \langle \textbf{v}_p |(\textbf{r}-\textbf{R}_1)|0_{\mathcal L} \rangle, \nonumber\\
&=& - \langle \textbf{v}_p |\textbf{r}|0_{\mathcal L} \rangle +\textbf{R}_1\langle \textbf{v}_p |0_{\mathcal L} \rangle. 
 \label{Eq:dL}
\end{eqnarray}
Similarly for ${b}_{\mathcal R}( \textbf{v},t)$ we define a bound-free transition dipole matrix, $\textbf{d}_{\mathcal R}( \textbf{v})=- \langle \textbf{v}_p |(\textbf{r}-\textbf{R}_2)|0_{\mathcal R} \rangle$, and the total bound-free transition dipole matrix is thus: $\textbf{d}_m( \textbf{v})=\textbf{d}_{\mathcal L}( \textbf{v})+\textbf{d}_{\mathcal R}( \textbf{v})$. For atomic systems the above analysis is not necessary since the atom is placed at the origin of the coordinates system. Furthermore, in the second term of Eq. (\ref{Eq:dL}) the continuum state $|{\bf v}\rangle$ is an eigenstate of the full atomic Hamiltonian $H_0$, therefore this extra term $\textbf{R}_1\langle \textbf{v}_p |0_{\mathcal L} \rangle$ disappears.

On the third term of Eq.~(\ref{Eq:Newbl}) we define the continuum-continuum transition matrix element $ \textbf{G}_m( \textbf{v}, \textbf{v}^{\prime} )=\langle \textbf{v} | \textbf{r} |\textbf{v}'\rangle$ that rely upon on the scattering states $|\textbf{v}\rangle$ and $|\textbf{v}'\rangle$ defined in Eq.~(\ref{Eq:SCS}) as:
\begin{equation}
\textbf{G}_m( \textbf{v}, \textbf{v}^{\prime} ) = \textit{i} \:\nabla_\textbf{v}\delta(\textbf{v}-\textbf{v}^{\prime}) - \textbf{R}_1\delta(\textbf{v}-\textbf{v}^{\prime}) + \textbf{g}_m (\textbf{v},\textbf{v}^{\prime}).
 \label{Eq:CC1m}
\end{equation}
The first term on the right-hand of Eq.~(\ref{Eq:CC1m}) describes the motion of a free electron in the continuum. It is associated to events where the laser-ionized electron is accelerated by the laser electric field without any probability of rescattering. The second one, the rescattering transition matrix element $\textbf{g}_m( \textbf{v}, \textbf{v}^{\prime} )$, accounts for all the rescattering processes concerning the entire molecule. For $\textbf{g}_m( \textbf{v}, \textbf{v}^{\prime} )$ the residual Coulomb potential has to be taken into account. In this sense it can be written as a sum of components representing each rescattering channel on the molecule. The second term of Eq.~(\ref{Eq:CC1m}) then reads as:
\begin{eqnarray}
\textbf{g}_m (\textbf{v},\textbf{v}^{\prime}) &=&\textbf{g}_{\mathcal{ LL}} (\textbf{v},\textbf{v}^{\prime}) +\textbf{g}_{\mathcal{RR}} (\textbf{v},\textbf{v}^{\prime}) + \textbf{g}_{\mathcal{ RL}} (\textbf{v},\textbf{v}^{\prime}) + \textbf{g}_{\mathcal{ LR}} (\textbf{v},\textbf{v}^{\prime}),\\
&=& \langle \textbf{v}_p | (\textbf{r} -\textbf{R}_1)|\delta\textbf{v}_{\mathcal L}'\rangle +\langle \delta\textbf{v}_{\mathcal R} | (\textbf{r} -\textbf{R}_2)|\textbf{v}_p'\rangle + \langle \delta\textbf{v}_{\mathcal L} | (\textbf{r} -\textbf{R}_1)|\textbf{v}_p'\rangle + \langle\textbf{v}_p | (\textbf{r} -\textbf{R}_2)|\delta\textbf{v}_{\mathcal R}'\rangle.\nonumber 
 \label{Eq:gm}
 \end{eqnarray}
The first two terms in the above equation contain information about spatially local processes involving only one of the atoms-the so-called Local terms. On the contrary, the last two ones are processes involving both atomic centres, henceforth we refer to them as Non-Local and Cross terms, respectively. 

In the next we include corrections on Eq.~(\ref{Eq:Newbl}), extending the analysis based on the orthogonality of the plane and rescattering waves, discussed before. The transition amplitude for the \textit{Left} states then reads as:
\begin{equation}
\begin{split}
 \dot{b}_{\mathcal L}( \textbf{v},t) =&  -i\left(\frac{\textbf{v}^2}{2}+ I_p -\textbf{R}_1 \cdot \textbf{E}(t) \right)\textit{b}_{\mathcal L}( \textbf{v},t) +i\textbf{E}(t) \cdot {\bf d}_{\mathcal L}({\bf v}) \\
 & +\,{\bf E}(t)\cdot\nabla_{\bf v} b_{\mathcal L}({\bf v},t)-i \textbf{E}(t)\cdot \int{\textit{d}^3 \textbf{v}^{\prime}\: \textit{b}_{\mathcal L}( \textbf{v}^{\prime},t)\:{\bf g}_m( \textbf{v}, \textbf{v}^{\prime})}.
 \label{Eq:BpuntoL}
\end{split}
\end{equation}
The transition amplitude for the \textit{Right} states can be found following exactly the same procedure, namely (i) projecting the entire Hamiltonian of the system on the \textit{Right} wavefunction Eq.~(\ref{Eq:PWavefR}) to get an equation similar to Eq.~(\ref{Eq:TempEq}); (ii) multiplying it by a scattering state, $\langle \textbf{v}'|$ and (iii) defining the bound-continuum and continuum-continuum transition matrix elements including their respective corrections. 

A general equation containing both of the processes mentioned before reads as
\begin{eqnarray}
\dot{b}_{j}( \textbf{v},t) &=&  -i\left(\frac{\textbf{v}^2}{2}+ I_p -\textbf{R}_{i} \cdot \textbf{E}(t) \right)\textit{b}_j( \textbf{v},t) +i\textbf{E}(t) \cdot {\bf d}_{j}({\bf v})\nonumber \\
& & +\,{\bf E}(t)\cdot\nabla_{\bf v} b_j({\bf v},t)-i \textbf{E}(t)\cdot \int{\textit{d}^3 \textbf{v}^{\prime}\: \textit{b}_j( \textbf{v}^{\prime},t)\:{\bf g}_m ( \textbf{v}, \textbf{v}^{\prime})},
 \label{Eq:New5}
 \end{eqnarray}
where the subscript $j$ represents either the \textit{Left}, -- $j={\mathcal L}$ or \textit{Right}, -- $j={\mathcal R}$ and $i=1,2$ is the position of the atom. For instance, to obtain the transition amplitude for the \textit{Left} states, Eq. (\ref{Eq:BpuntoL}), we need to set $j={\mathcal L}$ and $i=1$ in the above equation.

In the following, we shall describe how it is possible to compute the transition amplitude, $b_j({\bf v},t)$ by applying the zeroth and first order perturbation theory to the solution of the partial differential equation, Eq. (\ref{Eq:New5}).
According to the perturbation theory, the transition amplitude solution $b_j({\bf v},t)$ can be split into two parts: $b_{0,j}({\bf v},t)$ and $b_{1,j}({\bf v},t)$, i.e.~$b_j({\bf v},t)=b_{0,j}({\bf v},t)+b_{1,j}({\bf v},t)$.  The zeroth order solution $b_{0,j}({\bf v},t)$ and the first perturbative order solution $b_{1,j}({\bf v},t)$. These correspond to the direct and rescattering terms, respectively. As is known, the direct term describes the transition amplitude for a laser-ionized electron that never rescatters with the remaining molecular-ions. On the other hand, the rescattering term $b_{1,j}({\bf v},t)$ is referred to an electron that, once ionized in a particular center, has a certain probability of rescattering with each of the molecular-ions.

\subsection{Direct transition amplitude}

Let us consider the process where the electron is ionized from one of the atoms without probability to return to its parent ion. 
The last  two  terms in Eq.~(\ref{Eq:New5}) describes  the continuum-continuum transition, $\nabla_{\bf v} b_j({\bf v},t)$, without   the  influence of  the  scattering center, and $\int{\textit{d}^3 \textbf{v}^{\prime}\: \textit{b}_j( \textbf{v}^{\prime},t)\:{\bf g}_m ({\bf v},{\bf v}')}$ by  considering  the  core  potential. Here, ${\bf g}_m({\bf v},{\bf v}')$, denotes  the  rescattering  transition matrix element, where the potential core plays an essential role. 

As the direct ionization  process should have a larger  probability  compared  with the rescattering  one~\cite{Lewenstein1995, PRANoslen2015},  one might neglect the last term in Eq.~(\ref{Eq:New5}), $\textbf{g}_m(\textbf{v},\textbf{v}^{\prime})=\textbf{0}$. This is what we refer as zeroth order solution:
\begin{equation}
{\partial }_tb_{0,j}( \textbf{v},t) =-\textit{i}  \bigg(\frac{\textbf{v}^2}{2}+ I_p-\textbf{R}_{i} \cdot \textbf{E}(t) \bigg) \textit{b}_{0,j}( \textbf{v},t)+ \textit{i}\: \textbf{E}(t) \cdot  \textbf{d}_j( \textbf{v}) +\textbf{E}(t) \cdot \nabla_{\textbf{v}}\textit{b}_{0,j}( \textbf{v},t).
\label{Eq:dB1}
\end{equation}
The latter equation is easily solved by conventional integration methods (see e.g.~\cite{Lev}) and considering the Keldysh transformation~\cite{Ehlotzky1992,Keldysh1965}. Therefore, the solution is as follows:
\begin{equation}
\begin{split}
b_{0,j}( \textbf{p},t) =&\textit{i}   \:\int_0^t{\textit{d} \textit{t}^{\prime}\:\textbf{E}(t^{\prime})}\:\cdot \textbf{d}_{j}\left[ \textbf{p}+\textbf{A}(t^{\prime})\right] \\
&\times \exp\left(-\textit{i} \:\int_{t^{\prime}}^t{d{\tilde t} \: \left \{ [ \textbf{p}+\textbf{A}({\tilde t})]^2/2 +I_p -\textbf{R}_{i} \cdot  \textbf{E}({\tilde t}) \right\}}\right).
 \end{split}
 \label{Eq:b_0LR}
\end{equation}
Note that the above equation is written in terms of the canonical momentum ${\bf p}= {\bf v} - {\bf A}(t)$ ~\cite{Lewenstein1994}. Here, we have considered that the electron appears in the continuum with kinetic momentum ${\bf v}(t')={\bf v}-{\bf A}(t)+{\bf A}(t')$ at the time $t'$, where {\bf v}~is the final kinetic momentum (note that in atomic units ${\bf p}= {\bf v}$), and $\textbf{A}(t) =-\int^{t}{ \textbf{E}(t^{\prime})dt^{\prime}}$ is the associated vector potential.

Equation (\ref{Eq:b_0LR}) has a direct physical interpretation which is understood as the sum of all the ionization events that occur from the time $t'$ to $t$. Then, the instantaneous transition probability amplitude of an electron at a time $t'$, at which it appears into the continuum with momentum ${\bf v}(t')= \textbf{p}+\textbf{A}(t^{\prime})$, is defined by the argument of the $[0,t]$ integral in Eq.~(\ref{Eq:b_0LR}). Furthermore, the exponent phase factor denotes the ``semi-classical action", ${S}_{i}({\bf p},t,t^{\prime})$, that defines a possible electron trajectory  from the birth time $t'$, at position $\textbf{R}_{i}$, until the ``detection" one $t$ as:
 \begin{equation}
{S}_{i}({\bf p},t,t^{\prime}) = \int_{t^{\prime}}^{t}{\:d{\tilde t}\left\{[{\bf p}+\textbf{A}({\tilde t})]^2/2 +I_p -\textbf{R}_{i} \cdot  \textbf{E}({\tilde t})  \right \}}.
\label{Eq:S_R}
\end{equation}
Note that the transition amplitude equations obtained so far depend on the position from which the electron is tunnel ionized to the continuum. The semi-classical action ${S}_{i}({\bf p},t,t^{\prime})$ contains this dependency as well.

Considering we are interested in to obtain the transition amplitude $b_{0,j}({\bf p},t)$ at the end of the laser pulse, the time $t$ is set at $t=t_{\rm F}$. Consequently, we shall define the integration time window as: $t$:$\,\,[0,t_{\rm F}]$. Furthermore we set ${\bf E}(0) = {\bf E}(t_{\rm F}) = {\bf 0}$, in such a way to make sure that the laser electric field is a time  oscillating wave and  does not  contain static components (the  same arguments apply to the vector  potential  ${\bf A}(t)$). Finally, the total transition amplitude for the direct process taking place on our two-center molecular system reads as:
 \begin{eqnarray}
 b_0( \textbf{p},t)=  b_{0,{\mathcal L}}( \textbf{p},t) +  b_{0,{\mathcal R}}( \textbf{p},t).
 \label{Eq:b_0}
\end{eqnarray}

\subsection{Rescattering transition amplitude} 

In order to find the solution for the transition amplitude of the rescattered photoelectrons, $b_{1}({\bf v},t)$,  we have considered, in Eq.~(\ref{Eq:New5}), $\textbf{g}_m (\textbf{v},\textbf{v}^{\prime}) \not= \textbf{0}$.  The first-order solution, $b_{1}({\bf v},t)$, is then obtained by inserting the zeroth-order solution, $b_{0,j}({\bf p},t)$, in the right-hand side of Eq.~(\ref{Eq:New5}). Thereby we obtain a general equation to describe the rescattering process as:
 \begin{eqnarray}
\hspace{-0.55cm}\dot{b}_{1,jj'}( \textbf{v},t) &=&  -i\left(\frac{\textbf{v}^2}{2}+ I_p -\textbf{R}_{i} \cdot \textbf{E}(t) \right)\textit{b}_{1,j'j}( \textbf{v},t)-i \textbf{E}(t)\cdot \int{\textit{d}^3 \textbf{v}^{\prime}\: \textit{b}_{0,j}( \textbf{v}^{\prime},t)\:{\bf g}_{jj'} ( \textbf{v}, \textbf{v}^{\prime})}.
\label{Eq:b_1xy}
 \end{eqnarray}
Where $j$ denotes the atom from where the electron is released and $j'$ the one where the electron is rescattered. As the continuum-continuum rescattering matrix element is split in four terms, see Eq.~(\ref{Eq:gm}), the associated rescattering transition amplitude contains four terms as well, i.e.~:
\begin{eqnarray}
b_{1}({\bf v},t)=b_{1,{\mathcal{ LL}}}({\bf v},t) + b_{1,\mathcal{ LR}}({\bf v},t)  + b_{1,\mathcal{ RR}}({\bf v},t) +
+ b_{1,\mathcal{ RL}}({\bf v},t).
\label{Eq:b_11} 
 \end{eqnarray}
The above equation  contains information about all the possible rescattering scenarios which take place in our molecular system. In addition, a direct physical interpretation of each term can be inferred as following;
\begin{enumerate}[(1)]
\item The first term, $b_{1,\mathcal{ LL}}({\bf v},t)$, denotes electron-tunneling ionization from an atom located at $\textbf{R}_1$ and rescattering with the same parent ion. We refer this process as ``spatially localized'', since the electron performs a local-rescattering with the same atomic core $j=j'$ from which it was born. 

\item The process described by $b_{1,\mathcal{LR} }({\bf v},t)$ considers both atoms of the molecule. It represents an event where the electron is tunnel-ionized from an atom at $\textbf{R}_{1}$ and rescatters with the other atom at $\textbf{R}_{2}$. We call this process as ``Cross process''. In fact, there exists another process involving both atoms. It occurs when the electron is detached from an atom located at $\textbf{R}_{1}$ and rescatters with the same parent ion, but there is certain probability of electron emission from the other ion-core, placed at $\textbf{R}_2$. We label the latter as ``Non-Local process''. 

\item The other ``Local'' term, $b_{1,\mathcal{RR}}({\bf v},t)$ describes the same process as (1), but now for an atom located at $\textbf{R}_2$. 

\item Finally, $b_{1,\mathcal{RL}}({\bf v},t)$ represents the same process as in (2), but the tunnel-ionization process takes place at $\textbf{R}_2$.
\end{enumerate} 
The differential equation describing the local-rescattering processes is constructed considering $j=j'$. For processes localized at \textit{Left} we need to set $j=j'=\mathcal{L}$ and $i=1$ and for the ones at the \textit{Right} $j=j'=\mathcal{R}$ and $i=2$, respectively. In this way the transition amplitude for the Local processes read as:
 \begin{eqnarray}
\hspace{-0.6cm}b_{1,jj}( \textbf{p},t) &=&-\int_0^t{\textit{d}t^{\prime}}\int_0^{t^\prime}{\textit{d} \textit{t}^{{\prime}{\prime}}}\int{\textit{d}^3\textbf{p}^{\prime} \:\textbf{E}(t^{\prime})} \cdot \textbf{g}_{jj} \left[\textbf{p}+\textbf{A}(t^{\prime}),\textbf{p}^{\prime}+\textbf{A}(t^{\prime})\right] \exp{ \left[-\textit{i} S_i({\bf p},t,t') \right] } \nonumber\\ 
&& \times  \exp{\left[-\textit{i}  S_i({\bf p}',t',t'')\right]} \: \textbf{E}(t^{{\prime}{\prime}}) \cdot \textbf{d}_j \big[ \textbf{p}^{\prime} +\textbf{A}(t^{{\prime}{\prime}})\big].\label{Eq:b_1L2}
 \end{eqnarray}  
As we expect the rescattering transition amplitude contains two exponential factors, each representing the excursion of the electron in the continuum: before and after the rescattering event. In the above equation both phases factors needs to be evaluated with the same subscript, i.e.~$S_1({\bf p},t,t')-S_1({\bf p}',t',t'')$ or $S_2({\bf p},t,t')-S_2({\bf p}',t',t'')$ since they are Local processes.

The last factor in Eq.~(\ref{Eq:b_1L2}), $\exp{\left[-\textit{i} {S}_i({\bf p}^{\prime},t{'},t'')\right]}$, represents the accumulated phase of an electron born at the time $t^{\prime\prime}$ in $\textbf{R}_i$ until it rescatters at time $t^\prime$. In the same way $\exp\left[-\textit{i}{S}_i({\bf p},t,t')\right]$ defines the accumulated phase of the electron after it rescatterss at a time $t'$ to the ``final" one $t$,  when the electron is ``measured" at the detector with momentum \textbf{p}. Finally, the quantity $\textbf{E}(t^{{\prime}{\prime}}) \cdot \textbf{d}_j\left[\textbf{p}^{\prime} +\textbf{A}(t^{{\prime}{\prime}})\right]$ is the probability amplitude  of an emitted electron  at  the  time $t^{\prime\prime}$ that has a kinetic momentum of ${\bf v}^{\prime}(t'')=\textbf{p}^{\prime}+\textbf{A}(t^{{\prime}{\prime}})$. 
Similarly, to find the transition amplitude for a Local process at the \textit{Right} atom, we need to consider $j=j'={\mathcal R}$ and $i=2$ in Eq.~(\ref{Eq:b_1L2}) and use the \textit{Right} dipole transition matrix element. 

The Cross and Non-Local processes are formulated by considering $j\neq j'$ in the following way: $j={\mathcal L},{\mathcal R}$ , $j'={\mathcal R},{\mathcal L}$ in Eq.~(\ref{Eq:b_1xy}). The phase factors have to be set in different atomic positions, it means $S_1({\bf p},t,t')-S_2({\bf p}',t',t'')$ or $S_2({\bf p},t,t')-S_1({\bf p}',t',t'')$. For instance, the transition amplitude for the \textit{Left-Right} reads as:
\begin{eqnarray}
\hspace{-0.65cm}b_{1,{\mathcal {LR}}}( \textbf{p},t) &=& -\int_0^t{\textit{d}t^{\prime}}\int_0^{t^\prime}{\textit{d} \textit{t}^{{\prime}{\prime}}}\int{\textit{d}^3\textbf{p}^{\prime} \:\textbf{E}(t^{\prime})} \cdot \textbf{g}_{\mathcal {LR}} \left[\textbf{p}+\textbf{A}(t^{\prime}),\textbf{p}^{\prime}+\textbf{A}(t^{\prime})\right] \exp{\left[-\textit{i}   S_2({\bf p},t,t')\right]} \nonumber \\
&& \times \:\textbf{E}(t^{{\prime}{\prime}}) \cdot \textbf{d}_{\mathcal L}\big[ \textbf{p}^{\prime} +\textbf{A}(t^{{\prime}{\prime}})\big]  \exp{\left[-\textit{i}   S_1({\bf p}',t',t'')\right]}.
\label{Eq:b_1LR1}
 \end{eqnarray} 
Here we notice that the above equation describes the atomic system presented in Ref.~\cite{PRANoslen2015} when the internuclear distance goes to zero, ${\textbf R}\to 0$. The verification of this limit for the direct process is straightforward. The phase factor, Eq.~(\ref{Eq:S_R}), becomes the well known semi-classical action: $S({\bf p},t,t')$ and the transition amplitude exactly has  the same dependency as for an atom, if we replace the atomic matrix elements on it. For the rescattering events, we have to neglect in Eq.~(\ref{Eq:b_11}) the contribution of the Non-Local and Cross terms (the last two terms) and follow the same procedure as before. In the following sections we obtain the exact dependency of the rescattered matrix elements and demonstrate that the atomic limit can also be recovered when ${\textbf R}\to 0$.
 
In the total rescattering transition amplitude, Eq.~(\ref{Eq:b_11}), we can identify two main contributions, namely, one generated for the Local processes and the other one for the Non-Local and Cross processes. In this way we define the rescattering transition amplitude as:
 \begin{eqnarray}
b_{1}( \textbf{p},t)=b_{Local}( \textbf{p},t) +b_{Non-Local+Cross}( \textbf{p},t),
\label{Eq:b_1}
 \end{eqnarray} 
 where
  \begin{eqnarray}
b_{Local}( \textbf{p},t)=b_{1,\mathcal{ LL}}( \textbf{p},t) +b_{1,\mathcal{ RR}}( \textbf{p},t),
\label{Eq:b_Local}
 \end{eqnarray} 
 and 
 \begin{eqnarray}
b_{Non-Local+Cross}( \textbf{p},t)=b_{1,\mathcal{ LR}}( \textbf{p},t) +b_{1,\mathcal{ RL}}( \textbf{p},t).
\label{Eq:b_Nlocal}
 \end{eqnarray}   
The total photoelectron spectra, $|b({\bf p},t_{\rm F})|^2 $, is a coherent superposition of both the direct $b_0({\bf p},t_{\rm F})$ and rescattered $b_1({\bf p},t_{\rm F})$ transition amplitudes, i.e.
 \begin{eqnarray}
|b({\bf p},t_{\rm F})|^2 &=& |b_0({\bf p},t_{\rm F})+b_1({\bf p},t_{\rm F})|^2,\nonumber \\ 
&=& |b_0({\bf p},t_{\rm F})|^2 + |b_1({\bf p},t_{\rm F})|^2 + b_0({\bf p},t_{\rm F}){b_1^*}({\bf p},t_{\rm F}) + c.c.
 \label{Eq:b_T}
\end{eqnarray}
The direct transition amplitude, Eq.~(\ref{Eq:b_0LR}), is a ``single time integral", and can be
computed straightforwardly.  For the rescattering one, Eqs.~(\ref{Eq:b_1L2}) and (\ref{Eq:b_1LR1}), the multiple time (``2D") and momentum (``3D") integrals present a demanding task from a computational perspective. In order to reduce the computational difficulties, and to obtain a physical interpretation of the ATI process, we shall employ the stationary phase method to partially evaluate these highly oscillatory integrals.

The fast oscillations of the momentum ${\bf p}'$ integral, suggests the utilization of the stationary-phase approximation or saddle point method to solve it in Eq.~(\ref{Eq:b_1L2}). This method is expected to be accurate, when both the $U_p$ and the $I_p$, as well as
the involved momentum ${\bf v}$ and ${\bf v}'$, are large. 
The quasi-classical action for the two-center molecule model, Eq.~(\ref{Eq:S_R}), can be rewritten as:
 \begin{equation}
 {S}_{i}({\bf p}',t',t'') =  \textbf{R}_{i} \cdot  [ \textbf{A}(t') - \textbf{A}(t'')] + \int_{t^{\prime}}^{t}{\:d{\tilde t}\left\{[{\bf p}'+\textbf{A}({\tilde t})]^2/2 +I_p  \right \}},
\end{equation}
where ${S}({\bf p}',t',t'') =   \int_{t''}^{t'}{\:d{\tilde t}\left[({\bf p}' + \textbf{A}({\tilde t}))^2/2 +I_p\right]}$, is proportional to $I_p$, $U_p$ and ${\bf v}'^2$,  and the phase factor, $\exp[-\textit{iS}(\textbf{p}^{\prime},t',t'')]$, oscillates very
rapidly. Then, the integral over the momentum $\mathbf{p}^{\prime}$ of Eq.~(\ref{Eq:b_1L2}) tends towards  zero
except near the  extremal points  of the  phase, i.e.~when $\nabla_{{\bf p}^{\prime}} \textit{S}(\textbf{p}^{\prime})={\bf
0}$. Thus, the main contributions to the momentum integral are
dominated by momenta, ${\bf p}'_s$, which satisfy the solution of
the equation: $ \nabla_{{\bf p}^{\prime}}
\textit{S}(\textbf{p}^{\prime})|_{{\bf p}'_s}={\bf 0} $. 
These saddle point momenta read:
\begin{equation}
\textbf{p}'_s = -\frac{1}{\tau}\int_{t^{{\prime}{\prime}}}^{t^\prime}  { \textbf{A}(\tilde{t}) \textbf{d}\tilde{t}}.
\end{equation}
Here, $\tau=t^{\prime}-t^{\prime \prime}$ is the excursion time of
the electron in the continuum. In terms of Classical Mechanics,
these momenta roots ${\bf p}'_s$ are those corresponding to the
classical electron trajectories because the  momentum  gradient of
the  action  can be understood as the  displacement  of a
particle~\cite{Goldstein}. As the momentum gradient of the action
is null $\Delta {\bf r}=\nabla_{{\bf p }'}S({\bf p}',t',t'')={\bf
0}$, the considered electron trajectories, ${\bf r}(t)$, are for
an electron that is born at the time $t''$ at a certain position
${\bf r}(t'')={\bf r}_0$. Then, after some time $t'$ the electron
returns to the initial position ${\bf r}(t')={\bf r}_0$ with an
average momentum ${\bf p}'_s$.
Therefore, the function $ \textit{S}(\textbf{p}^{\prime},t',t'') $ can be expanded in a Taylor series around the roots $\textbf{p}'_s$ and then apply the standard saddle point method to the 3D momentum integral over ${\bf p}'$ in all the rescattering equations, $b_{1,jj'}( \textbf{p},t)$
\begin{eqnarray}
\label{sing1}
&&\int{\textit{d}^3\textbf{p}^{\prime}  \textit{f}\:(\textbf{p}^{\prime}) \exp{\left(-\textit{i} S(\textbf{p}^{\prime})\right)}}=\\
&&\int{\textit{d}^3\textbf{p}^{\prime}  \textit{f}\:(\textbf{p}_s^{\prime}) \exp{\left(-\textit{i}\left[S(\textbf{p}_s^{\prime})+\frac{1}{2}\nabla^2_{{\bf p}'}S({\bf p}')\biggr|_{{\bf p}'_s} \cdot ({\bf p}'-{\bf p}'_s)^2\right]\right)}}\nonumber\\
&&   \approx \left( \frac{\pi}{\varepsilon +\frac{\textit{i}(t'-t'')}{2}}  \right)^\frac{3}{2} \exp{\left(-\textit{i} S(\textbf{p}_s^{\prime})\right)} \textit{f}\:(\textbf{p}_s^{\prime}).\nonumber
 \end{eqnarray}
Here, we have introduced an infinitesimal  parameter, $\varepsilon$, to avoid the divergence at  $t'=t''$.  Still, the singularity is not integrable and practically impossible to be treated numerically. One should stress out, however, that it is the result of the saddle point approximation restricted exclusively to the classical action. We have regarded in the calculation that the function  $\textit{f}\:(\textbf{p}^{\prime})$ is localized at a certain scale and consequently the singularity would simply disappear. This observation and the simple method to handle it has been pioneered in Ref.~\cite{Lewenstein1994}, for more information see the previous discussion in Ref.~\cite{PRANoslen2015}. The simplest way to avoid the problem is to set  $\varepsilon$ small, but non-zero; throughout this paper we use $\varepsilon=0.4\ {\rm a.u.}=0.2/ I_p$. 

With the last equation, Eq.~(\ref{sing1}), we have substantially reduced the dimensionality of the problem, i.e.~from a 5D integral to a 2D one. This reduction is extremely advantageous from a computational viewpoint.  Moreover,  with  the  saddle  point method  a quasi-classical  picture  for the  rescattering transition amplitude  is obtained for molecular systems,  similarly to the atomic approach described in~\cite{Corkum1993, Lewenstein1995, PRANoslen2015}.

In order to calculate the total photoelectron spectra for the two-center molecular system, we first need to define the ground and the continuum states. After having found them we then compute the bound-free transition dipole matrix elements, ${\bf d}_{\mathcal L}({\bf v})$ and ${\bf d}_{\mathcal R}({\bf v})$, and  the continuum-continuum transition rescattering matrix element ${\bf g}_{m}({\bf v},{\bf v}')$. 
In the next section, we shall introduce a short-range potential model in order to analytically compute both the transition matrix elements and the final photoelectron momentum distribution.

\section{Above-threshold ionization in diatomic molecules}

\subsection{A simplified molecular model}

In this section we define a simplified molecular model to validate the general above described formulation and to compute the ATI photoelectron spectra. 
Let us consider a diatomic molecule constructed as two fixed nuclear centers under the SAE. We describe the interaction of the electron with each molecular nuclei by an non-local potential. The Hamiltonian $\hat{H}(\textbf{p},\textbf{p}^{\prime}) $ of the system in the momentum representation can be written as:
\begin{equation}
\label{Hm}
\hat{H}_{\textbf{M}}(\textbf{p},\textbf{p}^{\prime}) = \frac{\bf{p}^2}{2}\delta(\textbf{p}-\textbf{p}^{\prime}) + \hat{V}_{\textbf{M}}(\textbf{p},\textbf{p}^{\prime}).
\end{equation}
The first term on the right-hand side is the kinetic energy operator,  and the second one is the interacting non-local potential defined according to:
\begin{equation} 
\hat{V}_{\textbf{M}}(\textbf{p},\textbf{p}^{\prime}) =  -\gamma' \: \phi (\textbf{p}) \: \phi (\textbf{p}^{\prime}) \: e^{-\textit{i}\textbf{R}_2 \cdot (\textbf{p}-\textbf{p}^{\prime})}-\gamma' \:\phi (\textbf{p}) \: \phi (\textbf{p}^{\prime}) \: e^{-\textit{i}\textbf{R}_1 \cdot  (\textbf{p}-\textbf{p}^{\prime}) }.
\label{Eq:Vp}
\end{equation}
This potential describes the interaction between the active electron and each of the nuclei of the molecule, and depends on the internuclear relative vector position ${\bf R}=\textbf{R}_{2}-\textbf{R}_{1}$.
The function $\phi({\bf p})=\frac{1}{\sqrt{{\bf p}^2 +\Gamma^2}}$ is the same auxiliary function used in~\cite{Lewenstein1995, PRANoslen2015}. The parameters $\gamma'= \frac{\gamma}{2}$ and $\Gamma $ are constants related with the shape of the ground state (for more details see~\cite{PRANoslen2015}). 

\begin{figure}[htb]
            \subfigure[~SFA -Potential Energy Surface]{ \includegraphics [width=0.44\textwidth] {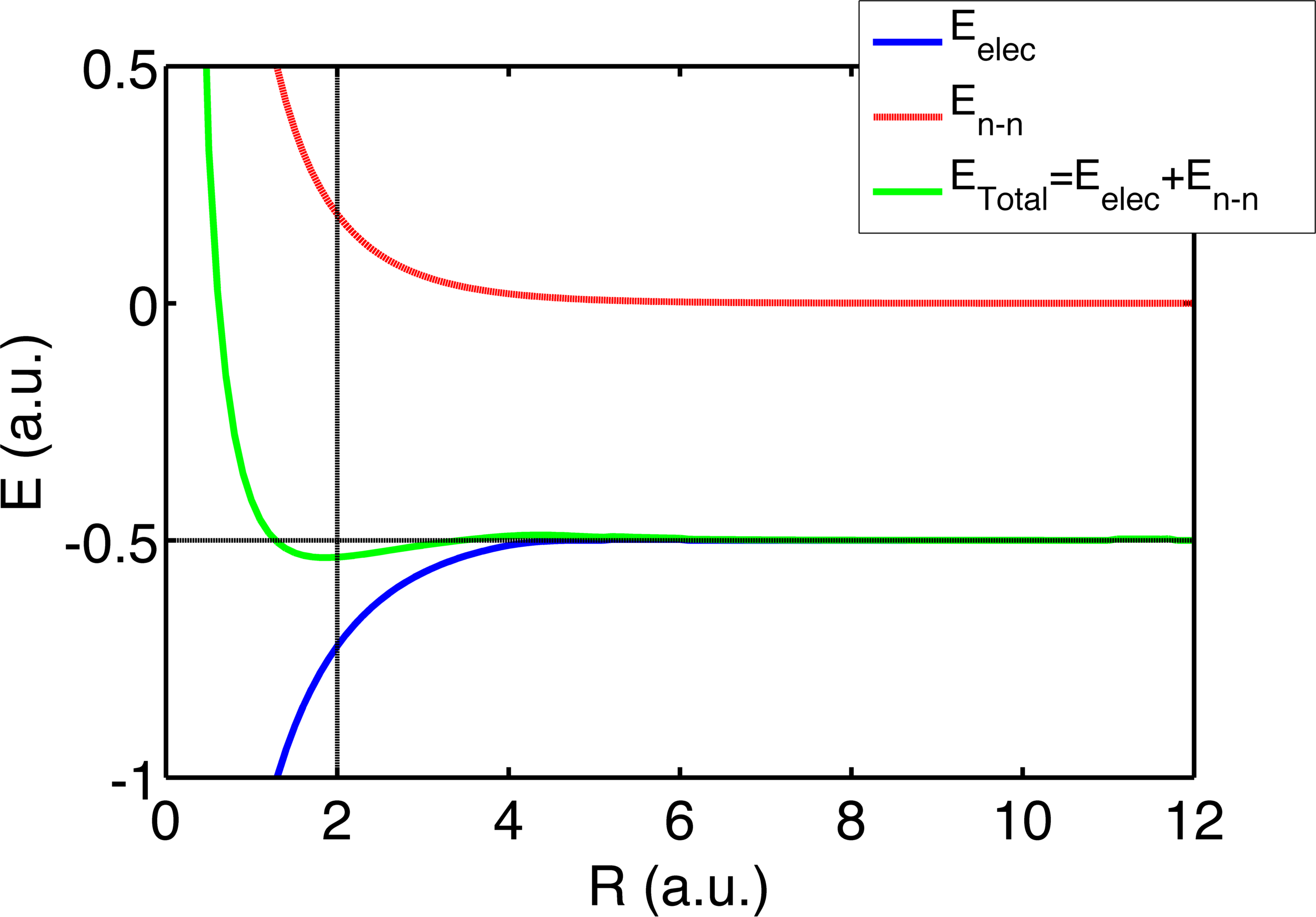}
            \label{fig:subfiga}}
                         \subfigure[~TDSE -Potential Energy Surface]{\includegraphics[width=0.44\textwidth]{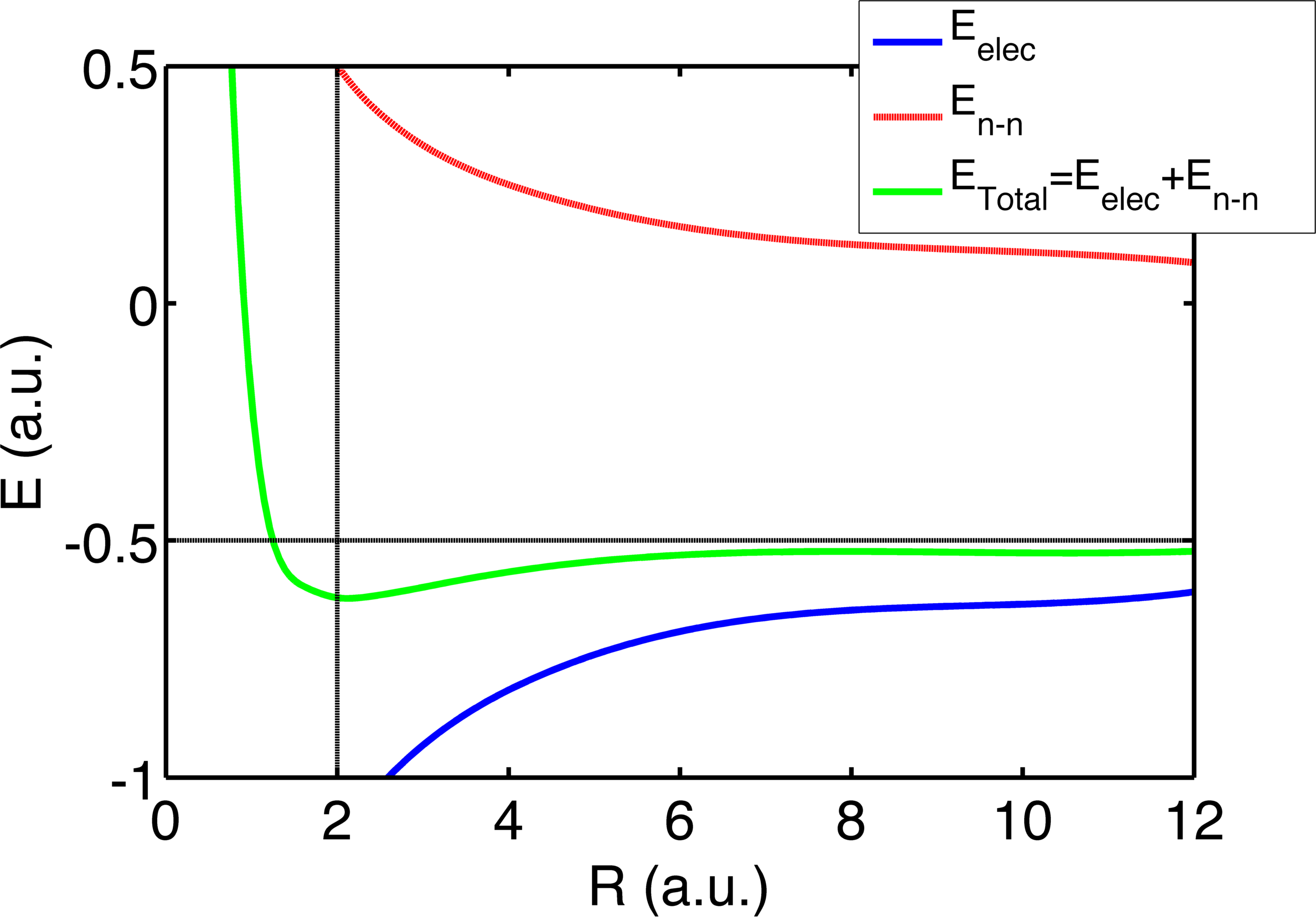}}
		    \caption{(Color online) Potential Energy Surface (PES) for the diatomic molecule H$_2^+$ as a function of the internuclear distance. (a) Electronic energy (blue line) calculated using Eq.~(\ref{Eq:Ee}), nuclear-nuclear energy (red line) and total energy of the system (green line) obtained with the SFA model. (b) The same as in (a) but computed by the numerical solution of the TDSE. The vertical dashed lines shows the energy minimum corresponding to the equilibrium distance of the system (see the text for details).}
		  	\label{Fig:PES}
		\end{figure}

By using $\hat{H}(\textbf{p},\textbf{p}^{\prime})$ from Eq.~(\ref{Hm}), we write the stationary  Schr\"odinger equation as follows:
\begin{eqnarray}
\hat{H}_{\textbf{M}}(\textbf{p},\textbf{p}^{\prime}) \Psi_{0\textbf{M}}(\textbf{p}) &=& \int{\textit{d}^3 \textbf{p}^{\prime}\hat{H}_{\textbf{M}}(\textbf{p},\textbf{p}^{\prime}) \Psi_{0\textbf{M}}(\textbf{p}')},\nonumber\\
\bigg(\frac{p^2}{2}  + I_p  \bigg) \Psi_{0\textbf{M}}(\textbf{p})&=& \gamma' \: \phi (\textbf{p}) \: e^{- \textit{i}\textbf{R}_{2} \cdot \textbf{p}}   \: \check{\varphi}_1+ \gamma' \: \phi (\textbf{p}) \: e^{-\textit{i}\textbf{R}_{1} \cdot \textbf{p} }   \: \check{\varphi}_2,
\label{Eq:Sch1}
\end{eqnarray}
where $I_p$ denotes the ionization potential energy of the wavefunction $\Psi_{0\textbf{M}}(\textbf{p})$ which is related to the ground potential energy by $E_{0} = -I_p$. To analytically solve Eq.~(\ref{Eq:Sch1}), in the momentum representation, we consider
 \begin{eqnarray}
\check{\varphi} _{1} &=&    \int{ \textit{d}^3 \textbf{p}^{\prime} \Psi_{0\textbf{M}}(\textbf{p}^{\prime}) \phi (\textbf{p}^{\prime}) e^{\textit{i}\textbf{R}_{2} \cdot \textbf{p}^{\prime}}} = \int{ \frac{ \textit{d}^3 \textbf{p}^{\prime}  \Psi_{0\textbf{M}}(\textbf{p}^{\prime})  e^{\textit{i}\textbf{R}_{2} \cdot \textbf{p}^{\prime} }}{\sqrt{  {p^{\prime}}^2 + \Gamma^2}}}, \label{Eq:gamma1}\\
\check{\varphi}_2 &=&  \int{ \textit{d}^3 \textbf{p}^{\prime} \: \Psi_{0\textbf{M}}(\textbf{p}^{\prime}) \phi (\textbf{p}^{\prime}) e^{\textit{i}\textbf{R}_{1} \cdot \textbf{p}^{\prime} }   } = \int{ \frac{ \textit{d}^3 \textbf{p}^{\prime}  \Psi_{0\textbf{M}}(\textbf{p}^{\prime})  e^{\textit{i}\textbf{R}_{1} \cdot \textbf{p}^{\prime}  }}{\sqrt{  {p^{\prime}}^2 + \Gamma^2}}},
\label{Eq:gamma2}
\end{eqnarray}
where the wavefunction for the bound states in momentum space $\Psi_{0\textbf{M}}(\textbf{p})$ is defined by 
 \begin{equation}
\Psi_{0\textbf{M}}(\textbf{p}) =  \frac{ \gamma' \: \check{\varphi}_1 \: e^{- \textit{i}\textbf{R}_{2} \cdot \textbf{p} }  }{\sqrt{(p^2 + \Gamma^2})(\frac{p^2}{2} + I_p)} +  \frac{\gamma' \: \check{\varphi}_2 \: e^{- \textit{i} \textbf{R}_{1} \cdot \textbf{p} }}{\sqrt{(p^2 + \Gamma^2})(\frac{p^2}{2} + I_p)}.
 \label{Eq:MWF1}
\end{equation}
Solving the system of Eqs.~(\ref{Eq:gamma1}) and (\ref{Eq:gamma2}), we find that, $\check{\varphi}_1 =\pm\check{\varphi}_2$. This relation let us two possible solutions, namely symmetric and anti-symmetric wavefunctions for $\Psi_{0\textbf{M}}({\bf p})$. Throughout this paper we shall only consider the symmetric wavefunction as follows:
 \begin{equation}
\Psi_{0\textbf{M}}(\textbf{p}) = \frac{ \mathcal{M} }{\sqrt{(p^2 + \Gamma^2})(\frac{p^2}{2} + I_p)} \bigg[2 \cos\bigg(\frac{\textbf{R} \cdot \textbf{p} }{2} \bigg) \bigg],
\label{Eq:Wf}
\end{equation} 
where $\mathcal{M}= \gamma' \: \check{\varphi}_1= \frac{\gamma}{2} \: \check{\varphi}_1$ is a normalization constant. This constant is obtained by employing the conventional normalization condition for the bound states. Consequently, this factor thereby reads
\begin{equation}
\mathcal{M}= \frac{1}{2}{\Bigg[ \frac{ 2\pi^2}{( 2I_p-\Gamma^2 )^2} \Bigg\{  \frac{2\: e^{-R \Gamma}} {R}  - \frac{2 \: e^{-R \sqrt{2I_p}}} {R} -\frac{(2I_p -\Gamma^2) e^{-R \sqrt{2I_p}}}{\sqrt{2I_p}} + \frac{(\sqrt{2I_p} - \Gamma)^2}{\sqrt{2I_p}}   \Bigg\} \Bigg] }^{-1/2}.
\label{Eq:BNormConst}
 \end{equation}
With the exact dependency of $\mathcal{M}$ we have well defined the bound state for our two-center molecular system. The wavefunction for the bound state can then be written as a combination of two \textit{Left} and \textit{Right} functions -$\Psi_{0\textbf{M}}(\textbf{p}) = \Psi_{0,{\mathcal L}}(\textbf{p}) + \Psi_{0,{\mathcal R}}(\textbf{p})$- in agreement with the photoelectron transition amplitude derivation 
\begin{equation}
\Psi_{0\textbf{M}}(\textbf{p}) =  \frac{ \mathcal{M} \: e^{-i\textbf{R}_1 \cdot  \textbf{p}} }{ \sqrt{(p^2 + \Gamma^2})(\frac{p^2}{2} + I_p)} + \frac{ \mathcal{M} \: e^{- i\textbf{R}_2 \cdot \textbf{p}}}{ \sqrt{(p^2 + \Gamma^2})(\frac{p^2}{2} + I_p)}.
\label{Eq:BSMolec}
\end{equation} 
In the above wavefunction we can clearly see that each term contains information about only one of the nuclei. The first term corresponds to the electron-wavefunction portion located at the atom on the \textit{Left} at $\textbf{R}_{1} = -\frac{\textbf{R}}{2}$, meanwhile the second one to the electron-portion placed on the \textit{Right} atom of the molecule at $\textbf{R}_{2} = +\frac{\textbf{R}}{2}$, respectively. 
Equations~(\ref{Eq:gamma1}) and (\ref{Eq:gamma2}) give us a relation between the electronic energy, $E_e$, of the molecular system and the internuclear distance $R$ as follows
 \begin{equation}
 \frac{ 2 \pi^2 \gamma} {R ( \Gamma^2-2E_e  )} \Bigg[e^{- \sqrt{2E_e} R}  - e^{-\Gamma R } + R (\Gamma - \sqrt{2E_e}) \Bigg] =1.
\label{Eq:Ee} 
 \end{equation}
In order to test the validity of the latter formulae in Fig.~\ref{Fig:PES} we show the Potential Energy Surface (PES) of the diatomic molecule, H$_2^+$, as a function of internuclear distance. We depict the different energy contributions, electronic and nuclear, of the molecular system H$_2^+$ obtained using the SFA model (left panel) and the exact solution of the TDSE (right panel). While the electron-nuclei interaction is described by a kind of non-local short-potential for our test molecular model, we choice as a repulsive potential between the nuclei a Yukawa one. 
We stress out that Fig.~\ref{Fig:PES} is in very good agreement with the PES reported in the literature \cite{H2plus}: it shows the minimum of equilibrium for an inter-atomic distance at $R_0=2$~a.u. This value is a clear signal of the good description offered by our SFA model. When $R$ is large the two atoms are weakly interacting and the energy of the system is equal to the energy of the atomic hydrogen, $-0.5$~a.u. As $R$ becomes smaller, the interaction results stronger and the energy is large and negative. In this case we say that a bond is formed between the atoms. At even smaller values of $R$, the internuclear repulsion is very large (red line), therefore the energy is large and positive. 

 \subsection{Bound-Continuum transition matrix element}

So far we have analytically obtained the ground state of our two-center molecular system. It allows us to compute the bound-free transition dipole matrix element, $\textbf{d}_{\mathcal L}( \textbf{p}_0)$ and $\textbf{d}_{\mathcal R}( \textbf{p}_0)$ by using Eq.~(\ref{Eq:dL}). By approximating the free or continuum  state as a plane wave with a given momentum, ${\bf p}_0$, the  bound-free  transition dipole matrix in the momentum representation reads
\begin{eqnarray}
\textbf{d}_{\mathcal L}( \textbf{p}_0) &&=-\textit{i}\nabla_{\textbf{p}}\Psi_{0,{\mathcal L}}(\textbf{p}) \Bigg\rvert_{{\bf p}_0} + \textbf{R}_1 \Psi_{0,{\mathcal L}}(\textbf{p}_0),\nonumber\\
&&=-2\textit{i}\: \mathcal{M}  \mathcal{A}( \textbf{p}_0) \: e^{-i\textbf{R}_1 \cdot \textbf{p}_0} ,
 \end{eqnarray}
for the atom on the \textit{Left}, meanwhile that for the one on the \textit{Right} it results: $\textbf{d}_{\mathcal R}( \textbf{v})=-2\textit{i}\: \mathcal{M}  \mathcal{A}( \textbf{p}_0) \: e^{-i\textbf{R}_{1} \cdot \textbf{p}_0}$. In both cases $\mathcal{A}( \textbf{p}_0)$ is defined as:
   \begin{equation}
   \mathcal{ A} ( \textbf{p}_0) =\frac{- \textbf{p}_0\: (3p_0^2 + 2I_p +2\Gamma^2)}{(p_0^2 + \Gamma^2)^{\frac{3}{2}}(p_0^2 + 2I_p)^2}.
   \end{equation}
\newline 
The  second important quantity to be calculated  before evaluating
the whole transition amplitude  $b({\bf p},t)$ is the transition continuum-continuum matrix  element, ${\bf g}_m({\bf p},{\bf p'})$.
Hence, we need to find the scattering  or continuum wavefunctions
of our model potential. Next, we shall calculate them by analytically  solving the time independent Schr\"odinger
equation  in the momentum  representation for positive energies.

\subsection{Scattering waves and the continuum-continuum transition matrix element}

Let us consider a scattering wave, $\Psi_{\textbf{M}{\bf p}_0}({\bf p})$, with asymptotic momentum ${\bf p}_0$, as a coherent superposition of a plane wave and an extra correction $\delta\Psi_{\textbf{M}{\bf p}_0}({\bf p})$
\begin{eqnarray}
\Psi_{\textbf{M}\textbf{p}_0}(\textbf{p})  = \delta(\textbf{p}-\textbf{p}_0) +  \delta\Psi_{\textbf{M}{\bf p}_0}(\textbf{p}).
\label{Eq.ScatState}
 \end{eqnarray}
 This state has an energy $E={{\bf p}_0^2}/2$. Then, the Schr\"odinger equation in momentum representation reads:
 \begin{equation}
\bigg(\frac{p^2}{2} - \frac{p_0^2}{2}  \bigg) \delta\Psi_{\textbf{M} \textbf{p}_0}(\textbf{p}) = -\hat{V}_{\textbf{M}}(\textbf{p},\textbf{p}_0)  
- \int{\textit{d}^3 \textbf{p}^{\prime} \hat{V}_{\textbf{M}}(\textbf{p},\textbf{p}^{\prime})\delta\Psi_{\textbf{M} \textbf{p}_0}(\textbf{p}^{\prime})}.
\label{Eq:ScSt1}
\end{equation}
Inserting the non-local potential, Eq~(\ref{Eq:Vp}), in Eq.~(\ref{Eq:ScSt1}) and after some algebra, we obtain:
\begin{equation}
\begin{split}
\big(p^2 - p_0^2 \big) \delta\Psi_{\textbf{M} \textbf{p}_0}(\textbf{p}) = & \:2  \gamma' \: \phi (\textbf{p}) \phi (\textbf{p}_0) \big[  e^{-\textit{i} \textbf{R}_{2} \cdot (\textbf{p} - \textbf{p}_0)} + e^{-\textit{i} \textbf{R}_{1} \cdot (\textbf{p} - \textbf{p}_0)}  \big]  \\ 
   &+ 2 \gamma' \check{\varphi}'_1\: \phi (\textbf{p})  e^{-i\textbf{R}_{2} \cdot  \textbf{p}}
 +  2\gamma' \check{\varphi}'_2\: \phi (\textbf{p})  e^{- i\textbf{R}_{1} \cdot  \textbf{p}},
\end{split}
\end{equation}  
where  the variables $\check{\varphi}'_1$ and $\check{\varphi}'_2$ are defined by:
 \begin{eqnarray}
\check{\varphi}'_1 &=&    \int{ \textit{d}^3 \textbf{p}^{\prime}  \delta\Psi_{\textbf{M} \textbf{p}_0} (\textbf{p}^{\prime}) \phi (\textbf{p}^{\prime}) e^{\textit{i}\textbf{R}_{2} \cdot \textbf{p}^{\prime} }} =  \int{\frac{ \textit{d}^3 \textbf{p}^{\prime}  \delta\Psi_{\textbf{M} \textbf{p}_0}(\textbf{p}^{\prime}) e^{\textit{i}\textbf{R}_{2} \cdot \textbf{p}^{\prime} }} {\sqrt{p'^2 + \Gamma^2}} },\\
\check{\varphi}'_2 &=& \int{ \textit{d}^3 \textbf{p}^{\prime}  \delta\Psi_{\textbf{M} \textbf{p}_0}(\textbf{p}^{\prime}) \phi (\textbf{p}^{\prime}) e^{\textit{i}\textbf{R}_{1} \cdot \textbf{p}^{\prime} }   } = \int{\frac{ \textit{d}^3 \textbf{p}^{\prime}  \delta\Psi_{\textbf{M} \textbf{p}_0}(\textbf{p}^{\prime}) e^{\textit{i}\textbf{R}_{1} \cdot \textbf{p}^{\prime} }} {\sqrt{p'^2 + \Gamma^2}} }.
\end{eqnarray}
Finally, for $\delta \Psi_{\textbf{M}{\bf p}_0}$ we write:
\begin{eqnarray}
 \delta\Psi_{\textbf{M} \textbf{p}_0}(\textbf{p}) & = &  \frac{ \mathcal{D}_1(\textbf{p}_0) \: e^{- i \textbf{R}_{2} \cdot (\textbf{p} - \textbf{p}_0) } - \mathcal{D}_2(\textbf{p}_0) \:e^{- i \textbf{R}_{2} \cdot (\textbf{p} + \textbf{p}_0) } }{ \sqrt{p^2 + \Gamma^2} \:(p_0^2 -p^2 + i \epsilon) }\nonumber \\
 && + \frac{ \mathcal{D}_1 (\textbf{p}_0) \: e^{-i \textbf{R}_{1} \cdot (\textbf{p} - \textbf{p}_0 )} - \mathcal{D}_2 (\textbf{p}_0) \: e^{-i \textbf{R}_{1} \cdot (\textbf{p} + \textbf{p}_0 )}}{ \sqrt{p^2 + \Gamma^2} \:(p_0^2 -p^2 +\textit{i}\epsilon) },
\label{Eq:RescM}
\end{eqnarray}
where $\epsilon$ is another infinitesimal parameter to avoid the divergence at the ``energy shell", $p^2=p^2_0$. The singularity at the ``energy shell" is avoided due to the finite spread of the involved wavepackets. 
In numerical calculations we set throughout this paper $\epsilon=0.4 \ {\rm a.u.}$ (for more details see~\cite{PRANoslen2015}).  
The integration ``constants" for the scattering states in Eq.~(\ref{Eq:RescM}) have the following dependency:
\begin{equation}
\mathcal{D}_1(\textbf{p}_0)=\frac{\gamma}{\sqrt{p_0^2 + \Gamma^2}} \Bigg\{  \frac{1 + I_1 }{ I_2^2 - \big(1+I_1  \big)^2  }  \Bigg\}; \mathcal{D}_2(\textbf{p}_0)=\frac{\gamma}{\sqrt{p_0^2 + \Gamma^2}} \Bigg\{  \frac{I_2 }{ I_2^2 - \big(1+I_1  \big)^2  }  \Bigg\},
\end{equation} 
where
\begin{eqnarray} 
I_1 &=&  \frac{-  2\pi^2 \: \gamma }{\Gamma - i\sqrt{|p_0^2 + \textit{i}\:\epsilon|}},\label{Eq:I1}\\
I_2 &=& \frac{- 2 \pi^2 \:\gamma} {R\:(p_0^2 + \Gamma^2  + i\epsilon)} \bigg[ e^{iR\:\sqrt{p_0^2 +i\epsilon}} - e^{-R\: \Gamma} \bigg ]\label{Eq:I2}.
\end{eqnarray}   
Finally, the molecular scattering wavefunction can be written as a composition of two contributions, namely 
  \begin{equation}
 \Psi_{\textbf{M} \textbf{p}_0}(\textbf{p})=  \delta(\textbf{p}-\textbf{p}_0) +\delta \Psi_{{\mathcal R} \textbf{p}_0}(\textbf{p}) + \delta\Psi_{{\mathcal L}\textbf{p}_0}(\textbf{p}),
 \end{equation}
 where, 
\begin{eqnarray} 
\delta\Psi_{{\mathcal L}\textbf{p}_0}(\textbf{p})= \frac{ \mathcal{D}_1 (\textbf{p}_0) \: e^{-i \textbf{R}_{1} \cdot (\textbf{p} - \textbf{p}_0 )} - \mathcal{D}_2 (\textbf{p}_0) \: e^{-i \textbf{R}_{1}  \cdot (\textbf{p} + \textbf{p}_0 )}}{ \sqrt{p^2 + \Gamma^2} \:(p_0^2 -p^2 +\textit{i}\epsilon) },
\label{Eq:dphiML} \\
 \delta\Psi_{{\mathcal R}\textbf{p}_0}(\textbf{p})= \frac{ \mathcal{D}_1 (\textbf{p}_0) \: e^{-i \textbf{R}_{2} \cdot (\textbf{p} - \textbf{p}_0 )} - \mathcal{D}_2 (\textbf{p}_0) \: e^{-i \textbf{R}_{2}  \cdot (\textbf{p} + \textbf{p}_0 )}}{ \sqrt{p^2 + \Gamma^2} \:(p_0^2 -p^2 +\textit{i}\epsilon) }.
 \label{Eq:dphiMR}
 \end{eqnarray}    
The Eq.~(\ref{Eq:dphiML}) describes electrons that has probability of scatter with the ion core placed at $\textbf{R}_{1}$. Similarly, Eq.~(\ref{Eq:dphiMR}) represents a scattering process with the nucleus placed at $\textbf{R}_{2}$.
 
Let us consider the scattering waves obtained in Eqs.~(\ref{Eq:dphiML}) and~(\ref{Eq:dphiMR}) to evaluate the continuum-continuum transition matrix element of Eq.~(\ref{Eq:gm}). After some algebra it reads as:
\begin{equation}
\textbf{g}_m(\textbf{p}_1, \textbf{p}_2) =  \mathcal{Q}_1(\textbf{p}_1, \textbf{p}_2)\: \Big[ e^{-i \textbf{R}_{1} \cdot (\textbf{p}_1 - \textbf{p}_2) }  + e^{- i \textbf{R}_{2} \cdot (\textbf{p}_1 - \textbf{p}_2) } \Big]+  \mathcal{Q}_2(\textbf{p}_1, \textbf{p}_2) \: \Big[ e^{- i \textbf{R}_{1} \cdot (\textbf{p}_1 + \textbf{p}_2) } + e^{-i \textbf{R}_{2} \cdot (\textbf{p}_1 + \textbf{p}_2) }\Big].
\label{Eq:gp1p2M}
\end{equation}
where
 \begin{eqnarray}
\mathcal{Q}_1 (\textbf{p}_1, \textbf{p}_2) &=& i \Big[ \mathcal{D}_1 (\textbf{p}_2) \mathcal{C}_1(\textbf{p}_1, \textbf{p}_2) - \mathcal{D}^*_1(\textbf{p}_1) \mathcal{C}_2(\textbf{p}_1, \textbf{p}_2) \Big],\\
\mathcal{Q}_2 (\textbf{p}_1, \textbf{p}_2) &=& -i \Big[ \mathcal{D}_2 (\textbf{p}_2) \mathcal{C}_1(\textbf{p}_1, \textbf{p}_2) -\mathcal{D}^*_2(\textbf{p}_1) \mathcal{C}_2(\textbf{p}_1, \textbf{p}_2) \Big],
\end{eqnarray}
and
\begin{equation}
\mathcal{C}_1 (\textbf{p}_1, \textbf{p}_2) =  \Bigg[  \frac{ \textbf{p}_1(3p^2_1 -p^2_2 + 2\Gamma^2 ) }{  (p_1^2 +\Gamma^2) ^{\frac{3}{2}}(p_2^2 - p_1^2 +i\epsilon)^2 }\Bigg],
\mathcal{C}_2(\textbf{p}_1, \textbf{p}_2) =  \Bigg[  \frac{\textbf{p}_2(3 p^2_2- p^2_1 + 2\Gamma^2 )}{  (p_2^2 +\Gamma^2)^{\frac{3}{2}} (p_1^2 - p_2^2 -\textit{i}\epsilon)^2 } \Bigg].
\end{equation}
From Eq.~(\ref{Eq:gp1p2M}) we can identify all the contributions, i.e.~Local, Non-Local and Cross as:
\begin{eqnarray}
\textbf{g}_{\mathcal L}(\textbf{p}_1, \textbf{p}_2) &=& \textbf{g}_{{\mathcal{LL}}}(\textbf{p}_1, \textbf{p}_2)  +\textbf{g}_{{\mathcal{LR}}}(\textbf{p}_1, \textbf{p}_2),\nonumber\\
\textbf{g}_{\mathcal L}(\textbf{p}_1, \textbf{p}_2) &=&  \mathcal{Q}_1(\textbf{p}_1, \textbf{p}_2) \: e^{- i \textbf{R}_{1} \cdot (\textbf{p}_1 - \textbf{p}_2) }   + \mathcal{Q}_2(\textbf{p}_1, \textbf{p}_2) \: e^{- i \textbf{R}_{1} \cdot (\textbf{p}_1 + \textbf{p}_2) },
\label{Eq:gLeft}
\end{eqnarray}
  and  
  \begin{eqnarray}
\textbf{g}_{\mathcal R}(\textbf{p}_1, \textbf{p}_2) &=& \textbf{g}_{\mathcal {RR}}(\textbf{p}_1, \textbf{p}_2)  +\textbf{g}_{\mathcal {RL}}(\textbf{p}_1, \textbf{p}_2),\nonumber\\
\textbf{g}_{\mathcal R}(\textbf{p}_1, \textbf{p}_2) &=& \mathcal{Q}_1(\textbf{p}_1, \textbf{p}_2)\: e^{- i \textbf{R}_{2} \cdot (\textbf{p}_1 - \textbf{p}_2) }  +  \mathcal{Q}_2(\textbf{p}_1, \textbf{p}_2) \:e^{ -i \textbf{R}_{2} \cdot (\textbf{p}_1 + \textbf{p}_2)}.
\label{Eq:gRight}
\end{eqnarray} 
After obtaining both the bound-free and continuum-continuum transition matrix elements is possible to compute the Eqs.~(\ref{Eq:b_0LR}) and (\ref{Eq:b_1L2}) to obtain the direct, the rescattering and the total photoelectron transition amplitudes. The presented model is an alternative way to describe the ATI process mediated by a strong laser pulse. Our two-center molecular model is an extension to the one presented in Ref.\cite{PRANoslen2015} and renders to the same atomic equations when $\textbf{R}$ is close to cero (see Appendix A for more details and proofs).

We stress out that the method is physically intuitive, and can be understood on the
basis of a quasi-classical picture, i.e.~electron trajectories.
This is the main difference of our approach in comparison to the
numerical solution of the TDSE, whose physical interpretation is,
in spite of its accuracy, frequently challenging. The main
advantage of the proposed model is that  Eqs.~(\ref{Eq:b_0LR}) and (\ref{Eq:b_1L2}) give a clear physical understanding of the ATI process and provide rich and useful information about both the laser field
and the diatomic molecular target, which are encoded into the complex
transition amplitude  $b({\bf p},t)=b_0({\bf p},t)+b_1({\bf
p},t)$. The  exact analytical  solutions of both the direct  and
rescattering  transition amplitudes are, however, not trivial
to obtain if no approximations are considered.  In particular, for
the rescattering photoelectrons,  the solution is even more
complex and depends, generally, of the laser electric field shape.


\section{Results and discussion}

Along this section we shall compare the outcomes of our new model for the ATI spectra emitted from a H$_2^+$ system to the exact numerical solution of the 3D-TDSE. A scan on different internuclear distances of the ionization probability and the whole momentum distribution along the polarization laser and molecular orientation axis shows that our model works reasonable well. Furthermore, split of contributions coming from the left and right nuclei and local, cross and non-local rescattering processes helps to distinguish which part of the photoelectron spectra is relevant for each kind of event. The molecular internuclear distance is retrieved probing that our model is capable to capture the structural information encoded on the photoelectron spectra. 

Finally, experimental photoelectron spectra on the O$_2^+$ molecule driven by a mid-IR source (3.1~$\mu m$) demonstrate that our simplified model is able to render the main physics behind the re-scattering process in a 'complex' symmetric diatomic molecule. 

\subsection{Comparison of SFA and TDSE models}
The numerical integration for the photoelectron spectra computation by means of Eqs.~(\ref{Eq:b_0LR}) and (\ref{Eq:b_1L2}) has been performed via a rectangular rule with particular emphasis on the convergence of the results.  As the final momentum
distribution, Eq.~(\ref{Eq:b_T}), is ``locally" independent of the
momentum ${\bf p}$, i.e.~$|b({\bf p},t)|^2$ can be computed
concurrently for a given set of ${\bf p}$ values. We have
optimized the calculation of the whole transition amplitude,
$|b({\bf p},t)|^2$, by using the OpenMP parallel
package~\cite{OpenMPBook} and the MPI paradigm~\cite{open_mpi}. The final momentum photoelectron
distribution, $|b({\bf p},t)|^2$, is computed both in a
1D-momentum line along~$p_z$, and in a 2D-momentum plane
$(p_y,\,p_z)$. We shall compare these results with the numerical
solution of the TDSE. We fix the parameters of the non-local potential to $\Gamma=1.0$
and $\gamma=0.1$~a.u. Such values describe the potential energy surface of Fig.~\ref{Fig:PES}, which is in good agreement with the expected energy dependency of the H$_2^+$ molecular system. 
We use in our simulations an ultrashort
laser pulse with central frequency $\omega_0=0.057$~a.u.
(wavelength $\lambda=800$ nm, photon energy, $1.55$~eV), with a
$\sin^2$ envelope shape with $N_c=4$ total cycles (this
corresponds to a full-width at half-maximum FWHM~$=5.2$ fs) and a
CEP $\phi_{0}=0$~rad. The time step is fixed to $\delta t=
0.02$~a.u., and the numerical integration time window is
$t$:~$[0,t_{\rm F}]$, where $t_{\rm F}= N_cT_0\approx 11$ fs and
$T_0=2\pi/\omega_0$ denote the final ``detection" time and the
cycle period of the laser field, respectively.

In addition, we perform the numerical integration of the 3D-TDSE by using the Crank-Nicolson algorithm in cylindrical coordinates $(\rho,z)$ where the polar angle $\varphi$ is neglected. This is well justified by considering the laser field is linearly polarized along the molecular $z$-axis and the fact that the magnetic momentum electron-quantum number $m$ remains as a conserved quantity during the whole evolution of the system. 
Thereby, the electronic Hamiltonian of our systems is $\hat{H}=\frac{\hat{p}_\rho^2}{2}+ \frac{\hat{p}_z^2}{2} + \hat{V}(\rho,z) +zE(t)$.
For the present numerical solution of the TDSE, we have fixed the
position grid step to $\delta\rho=\delta z = 0.2$~a.u., with a total number
of points for the $\rho-$axis of ${\rm N}_{\rho}=6000$ and the $z-$axis of ${\rm N}_z = 12 000$, respectively. The ground state  is computed via imaginary  time propagation with a time step of~$\delta
t=-0.02\,i$ and the Coulomb potential for our two-center molecule is given by: $V(\rho,z)= - \frac{1}{\sqrt{ \rho^2+(z+R/2)^2}}- \frac{1}{\sqrt{ \rho^2+(z-R/2)^2}}$. 
The strong-field laser-molecule interaction is simulated by evolving
the electronic ground state wavefunction in real time, with a time step of
$\delta t=0.02$~a.u., and under the action of both the molecular 
potential  and the laser electric field. 
At the end of the laser pulse $t_{\rm F}$, when the laser electric field is zero, we compute the final
photoelectron energy-momentum distribution $|b_{\rm TDSE}(p_\rho,p_z,t_{\rm F})|^2$, by projecting the ``free" electron wave packet, $\Psi_c(\rho,z,t_{\rm F})$, over plane waves. The wavepacket
$\Psi_c(\rho,z,t_{\rm F})$, is calculated by smoothly masking the bound
states from the entire wavefunction $\Psi(\rho,z,t_{\rm F})$.

\begin{figure}[htb]
            \subfigure[$\,$Internuclear distance scan.]{ \includegraphics [width=0.45\textwidth] {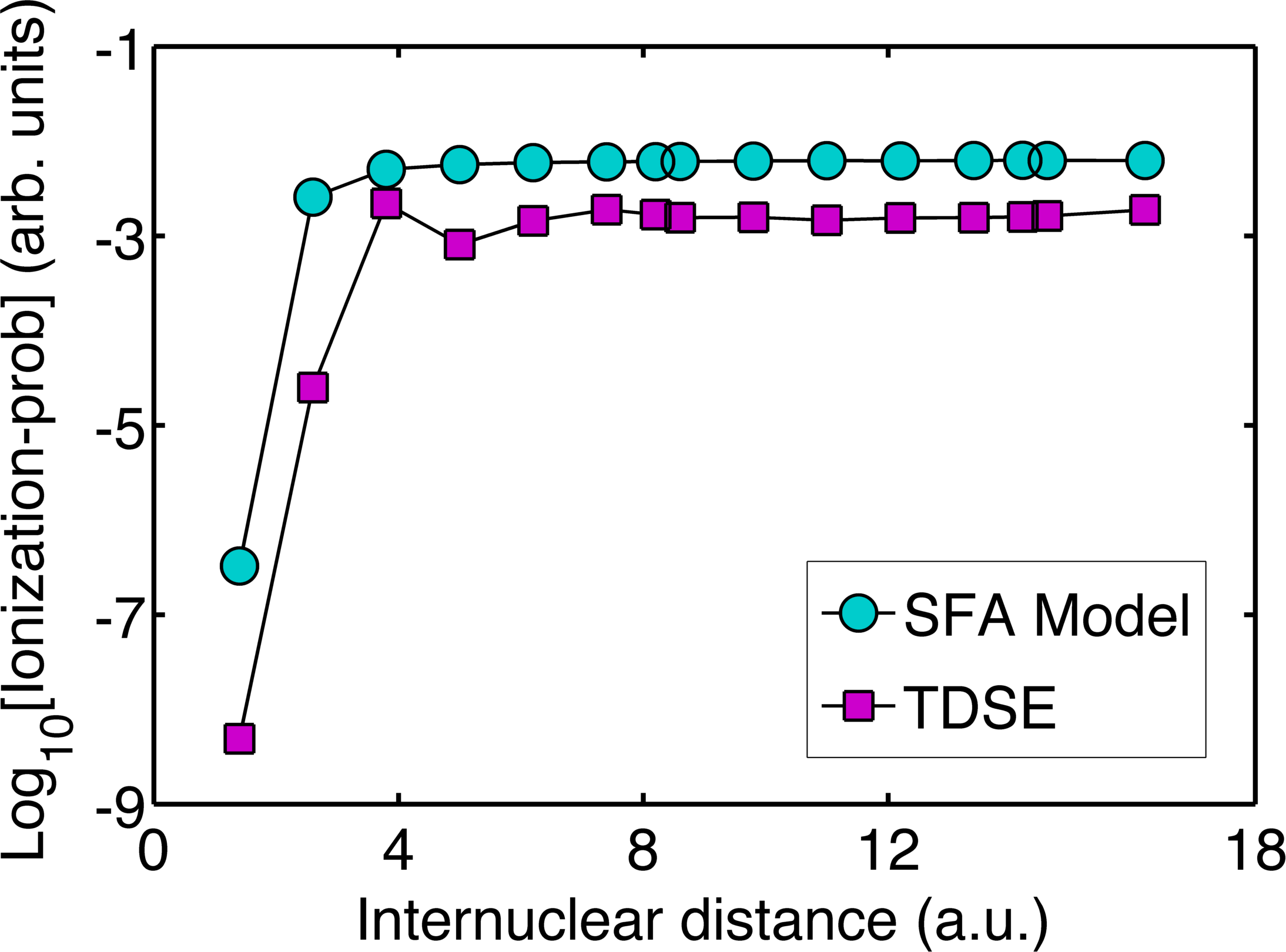}
            \label{fig:subfiga}}
                         \subfigure[$\,$SFA vs. TDSE at $R = 3.8$ a.u.]{\includegraphics[width=0.45\textwidth]{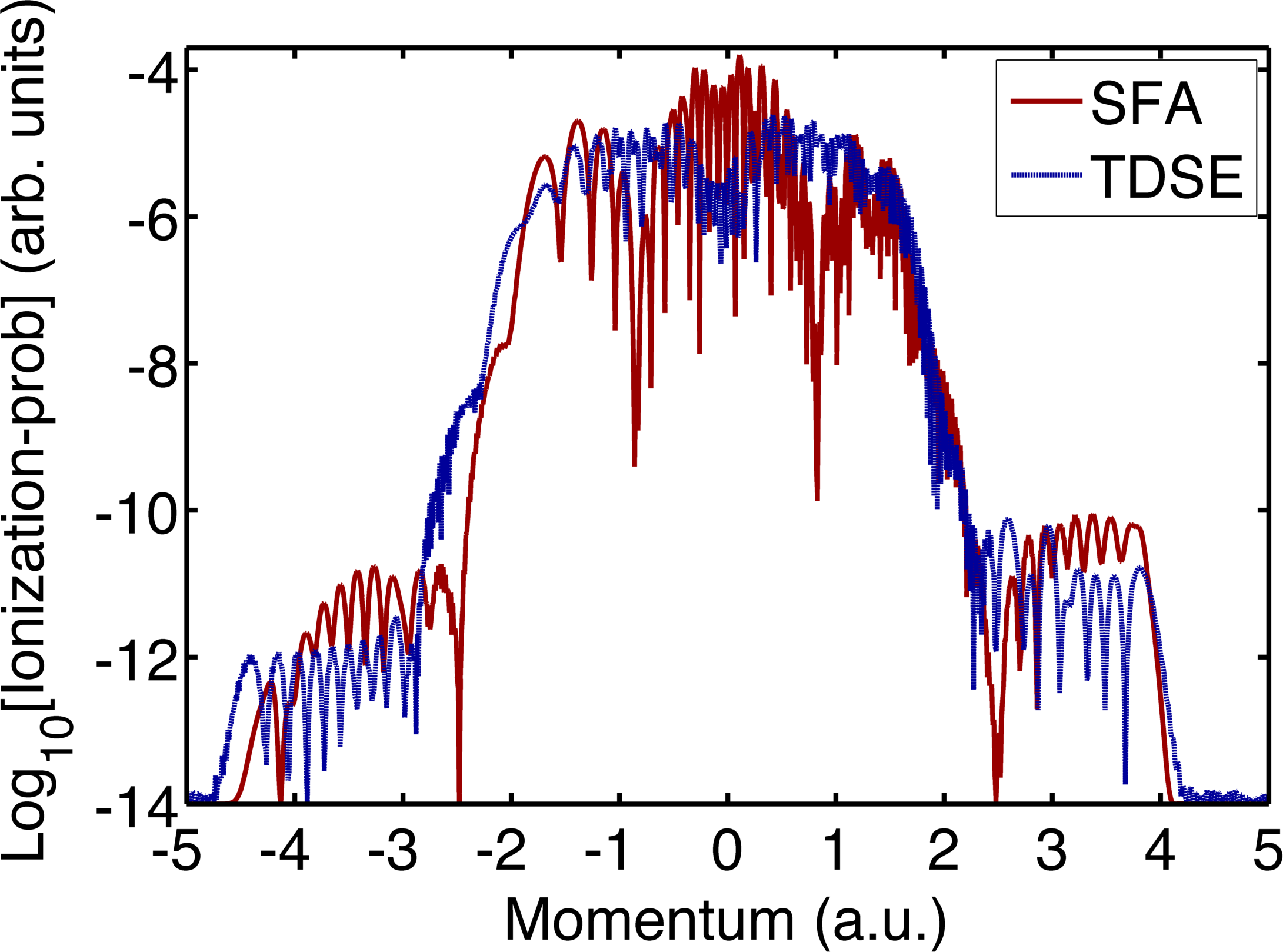}}
                                 
                         
                    \caption{(color online) (a) Ionization probability (in logarithmic scale) as a function of the internuclear distance $R$ calculated using the SFA (circle magenta) and the TDSE (square light blue) for $I_0= 4\times10^{14}$~W$\,\cdot$\,cm$^{-2}$. (b) Full transition amplitude $|b({p_z},t_{\rm F})|^2$ as a function of the photoelectron final momentum, calculated using the SFA model (red line) and ATI computed by the numerical solution of the TDSE (blue line) (see the text for details).
                }                  \label{Fig:SFA_TDSE}
\end{figure}

Firstly, and in order to test if our model is capable to capture the final photoelectron spectrum of the ATI processes, we compare our SFA model to the numerical solution of the 3D-TDSE for the simplified case of H$_2^+$.  Figure~\ref{Fig:SFA_TDSE} depicts such comparison. In Fig.~\ref{Fig:SFA_TDSE}(a) we calculate a scan of the ionization probability over a set of 15 interatomic distances. Note, that by ionization probability the reader should understand the final-time integral momentum distribution of the whole transition probability amplitude Eq.~(\ref{Eq:b_T}). Here we set the molecular axis parallel to the laser electric field polarization. Those results show a reasonable agreement between the SFA and TDSE models, particularly for larger internuclear distances, when the details of the potential are not important. In both calculations we can observe that for shorter distances the ionization probability is strongly dependent on the relative position of the atoms inside the molecule. The ionization probability scales almost exponentially (note that the scale is logarithmic) and increases rapidly, when the atoms are close each other, at $R\lesssim4$~a.u. Both models present the same trend, namely a low ionization probability for shorter distances, followed by a rapid increasing, and a sort of  `stabilization" for larger internuclear distances. The physical picture of this behavior is as follows: the electron is tightly (loosely) bounded for small (larger) internuclear distances. According to the Keldysh-Faisal-Reiss model the electrons have less probability to be ionized by tunneling effect~\cite{Reiss1980,Faisal1973,Keldysh1965} for potentials with larger $I_p$, which of course in this molecular case corresponds to small internuclear distances. Note that we mean small or large internuclear distances $R$ in comparison to the equilibrium one $R_0$.  Furthermore, the same tendency of both the SFA and the 3D-TDSE in the whole internuclear distances range is observed, except a constant factor, which clearly indicates the difference between the short-range (SFA) and long-range (TDSE) potentials. Further, the ionization probability shows a stabilization value (around $10^{-3}$ arb. units for both cases) from which it remains constant regardless the value of $R$. 

The previous comparison only describes the final photoelectron spectra dependence on the internuclear distance. A better scenario to evaluate the quality of our model, however, can be employed, namely, a one to one analysis of  the ATI momentum distributions. The aim is to confirm if our model is able to capture both the interference nature of the ATI spectra for molecules and the underlying electron dynamics.  In Fig.~\ref{Fig:SFA_TDSE}(b) we show results of the photoelectron momentum spectra computed by our quasi-classical model and the TDSE along the momentum line, $\textbf{p} = ({0}, {0}, {p}_z)$, at $R=3.8$ a.u. As in Fig.~\ref{Fig:SFA_TDSE}(a) we observe an excellent agreement between both models. It means that our quasi-classical approach is able to provide a reasonable good description of the whole ATI processes. We can argue that the two models are describing the same physics: stronger oscillations for small values of momentum followed by a rapid decrease of the ATI yield (at $|{p}_z|\lesssim2.5$ a.u.), a plateau, where the amplitude remains almost constant, and both approaches end up with an abrupt cutoff around the same value of $|{p}_z|\lesssim 4$~a.u.

\begin{figure}[htb]
 
            \subfigure[$\,$Total Contributions]{ \includegraphics [width=0.45\textwidth] {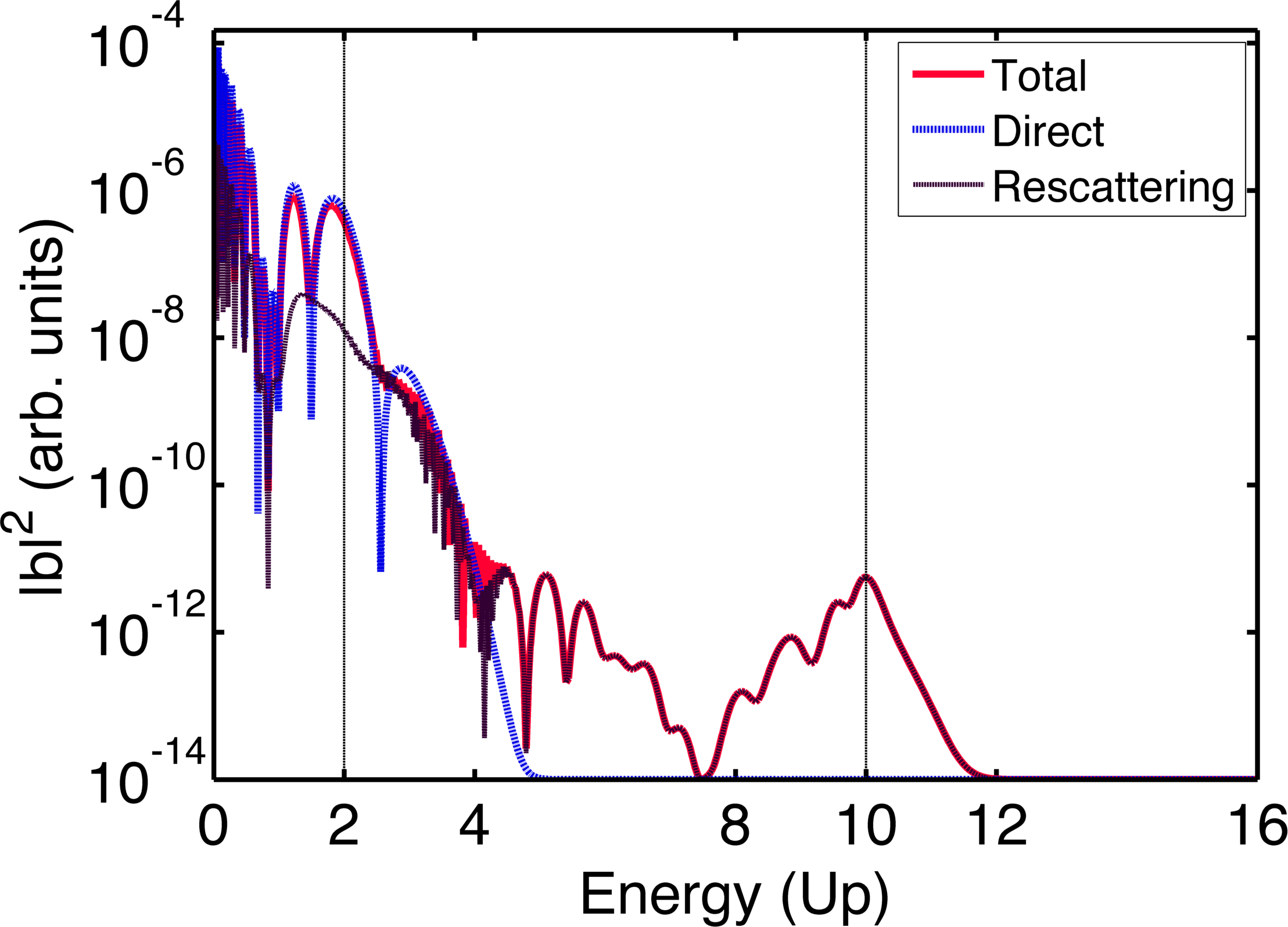}
            \label{fig:subfiga}}
		 \subfigure[$\,$Direct Contributions]{\includegraphics[width=0.45\textwidth]{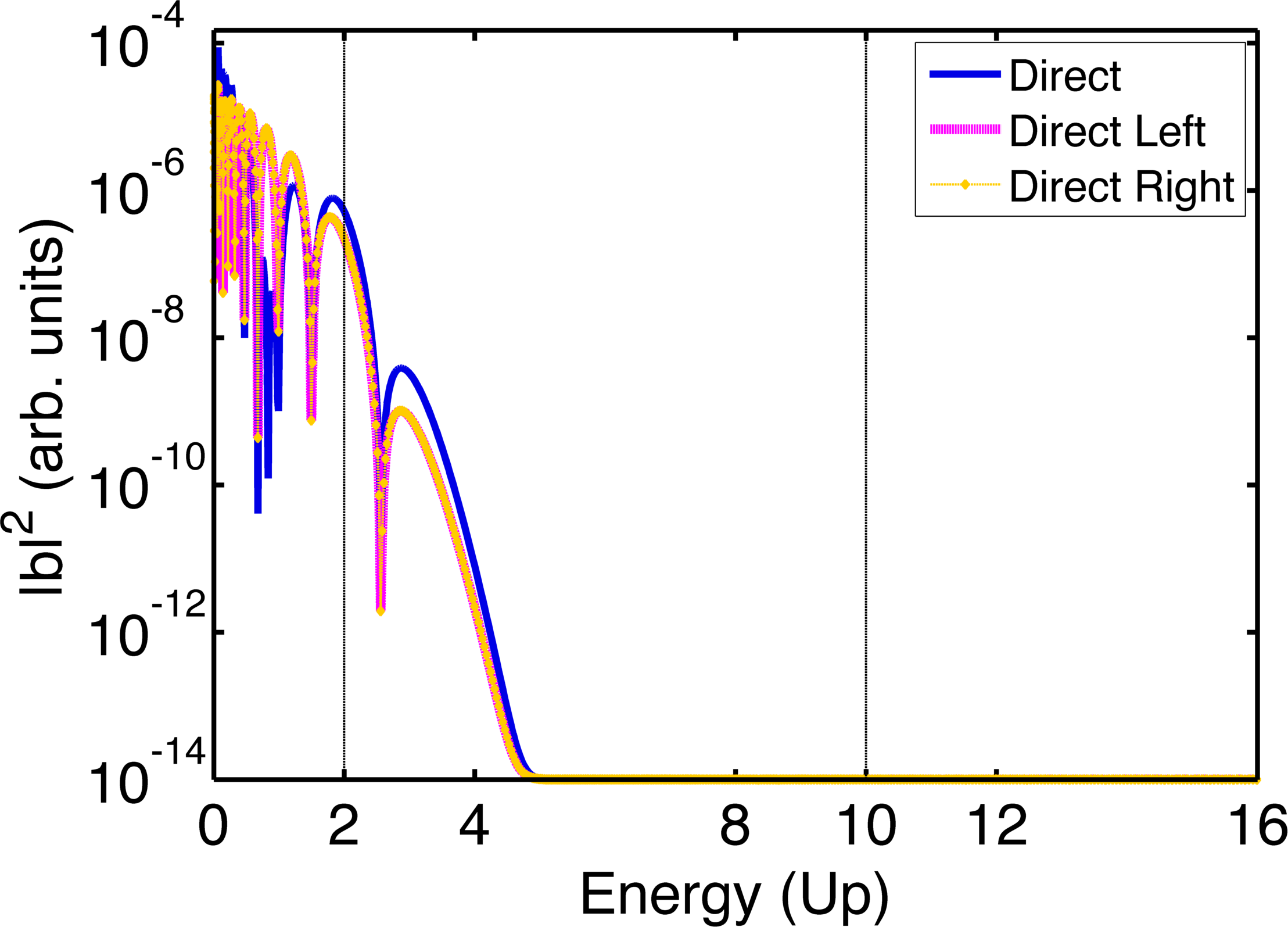}}
      		 \subfigure[$\,$Scattering Contributions]{\includegraphics[width=0.45\textwidth]{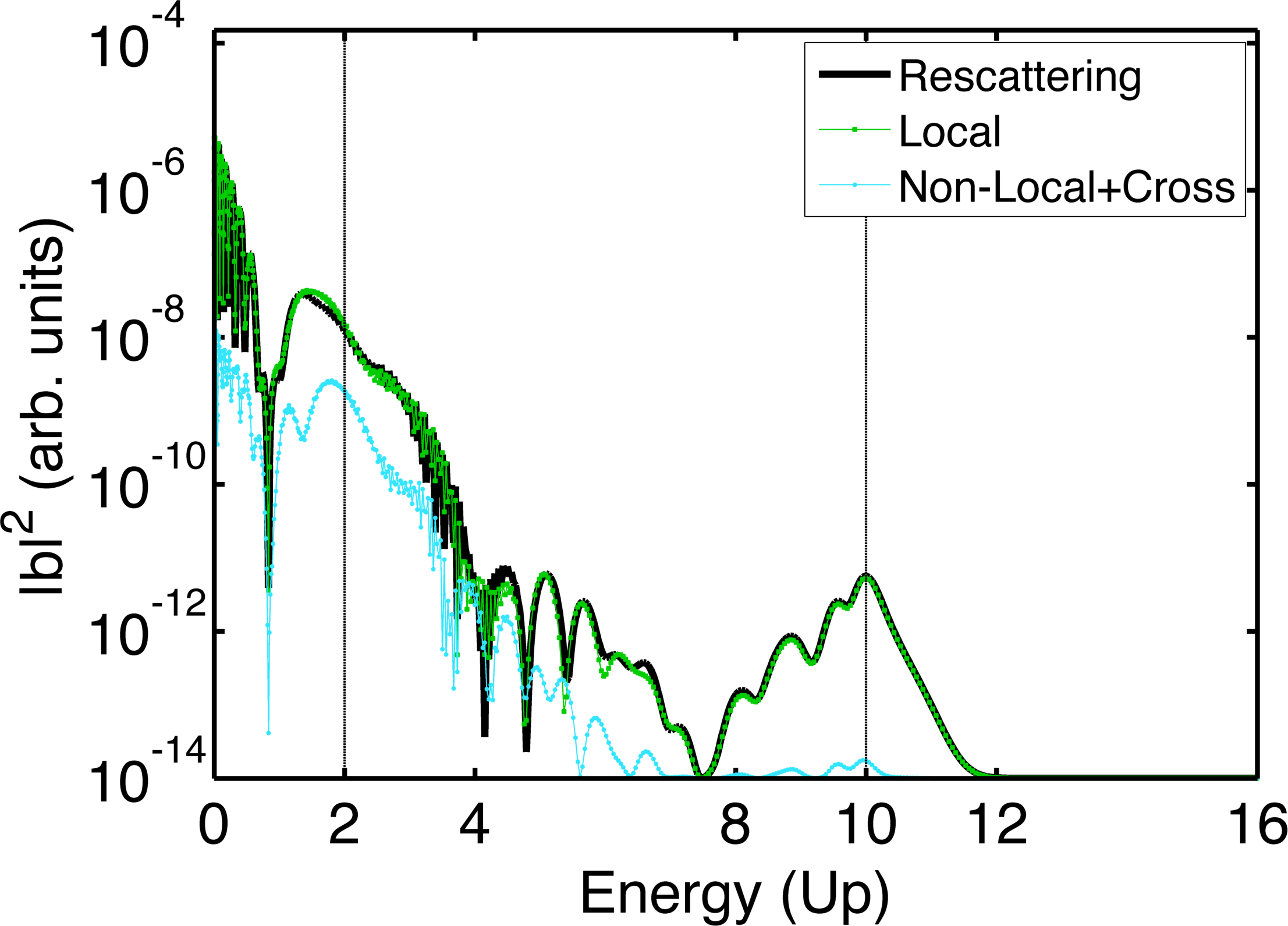}}
		 \subfigure[$\,$Local Contributions]{\includegraphics[width=0.45\textwidth]{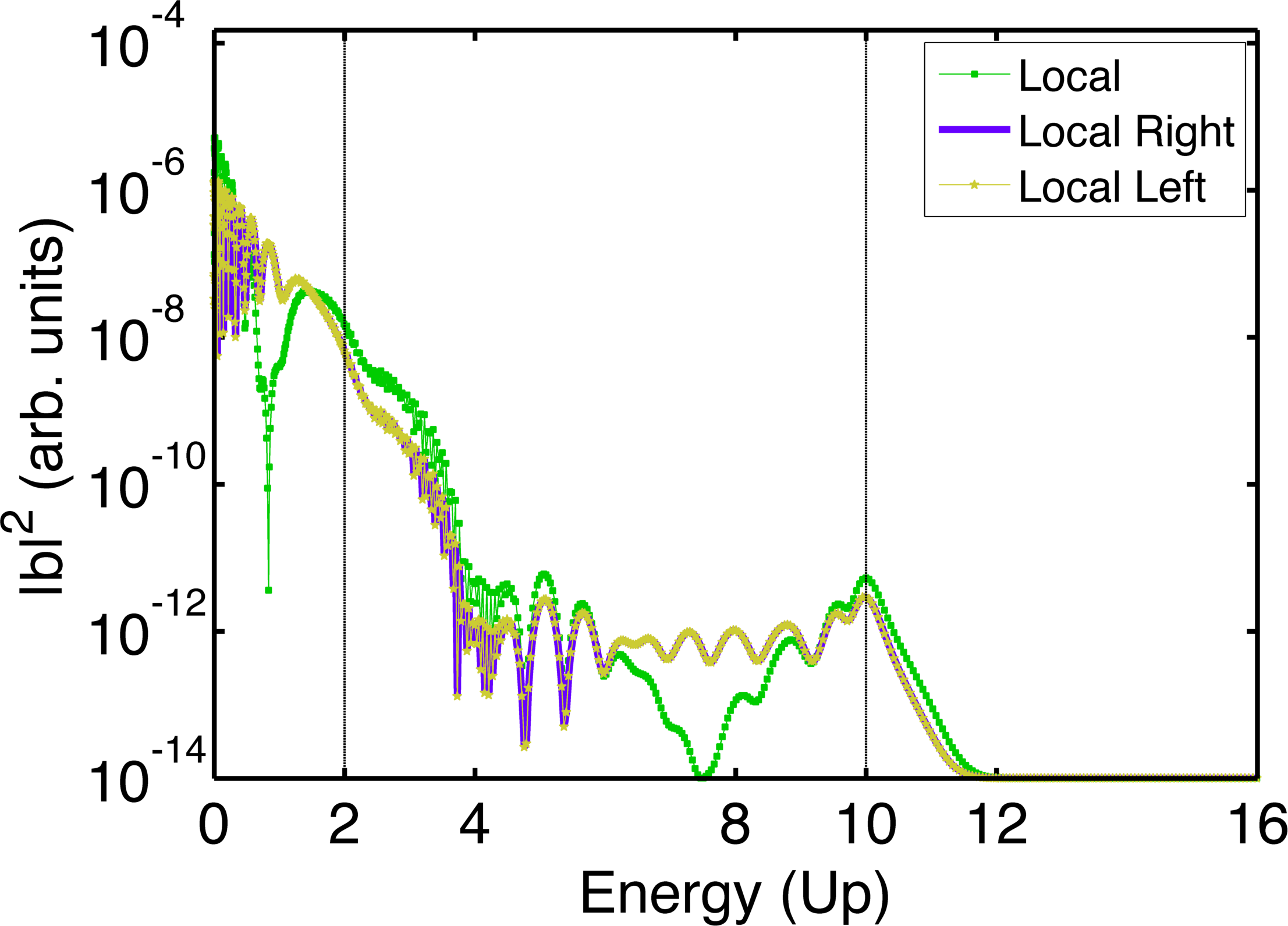}}
                         
             \caption{(color online) Total, direct and rescattering contributions of the photoelectron spectra (in logarithmic scale) as a function of the ponderomotive electron energy $U_p$ calculated by using our quasi-classical SFA model for $R=2.6$ a.u. The peak laser intensity used in this calculation is set to $I_0= 4\times10^{14}$~W$\,\cdot$\,cm$^{-2}$ (see text for more details).
                                    }                  \label{Fig:SFAContributions}                
                    \end{figure}

One of the main advantages of our SFA model is the possibility to disentangle the different contributions to the final ATI spectra (for details see previous Sections). In Fig.~\ref{Fig:SFAContributions} we show the different contributions, in logarithmic scale, as a function of the ponderomotive energy, $U_p$, for electrons with negative momenta along the $p_z$-direction and for a fixed value of ${R}$, close the equilibrium distance $R_0=2.0$~a.u. Figure~\ref{Fig:SFAContributions}(a) shows the main contributions to the full final photoelectron spectra: the total $|b(\textbf{p},t)|^2$, Eq.~(\ref{Eq:b_T}), the direct $|b_0(\textbf{p},t)|^2$, Eq.~(\ref{Eq:b_0}) and the rescattering $|b_1(\textbf{p},t)|^2$, Eq.~(\ref{Eq:b_1}) terms, respectively. In the same way in Fig.~\ref{Fig:SFAContributions}(b) we plot the two terms, $|b_{0,{\mathcal L}}(\textbf{p},t)|^2$ and $|b_{0,{\mathcal R}}(\textbf{p},t)|^2$ which contribute to the direct process. The terms that play an important role in the rescattering process, $|b_{Local}(\textbf{p},t)|^2$, Eq.~(\ref{Eq:b_Local}), and $|b_{Nl+Cross}(\textbf{p},t)|^2$, Eq.~(\ref{Eq:b_Nlocal}), are displayed in Fig.~\ref{Fig:SFAContributions}(c). As we can infer from the latter figure the main contribution to the rescattering term is from the Local processes (see Section II). Finally, in Fig.~\ref{Fig:SFAContributions}(d) we show the two processes contributing to the Local one. For this calculation we have considered the molecule aligned in the same direction as the laser electric field polarization, i.e.~the internuclear distance vector has only a $z$-component, $\textbf{R} = ({0}, {0}, {R}_z)$.

%
                    
Our second clear observation in Fig.~\ref{Fig:SFAContributions} is that each term contributes to different regions of the photoelectron spectra, i.e.~for electron energies $E_{p}\lesssim4 U_p$ the direct term $|b_0({\bf p},t)|^2$ dominates the spectrum and, on the contrary, it is the rescattering term, $|b_1({\bf p},t)|^2$ the one that prevails in the high-energy electron region. The photoelectron spectra show the expected two cutoffs defined by $2U_p$ and $10U_p$ (black dashed lines) which are presented in the atomic and molecular ATI process~\cite{Lewenstein1995,Milosevic2006}. As a consequence of this last observation we can safely argue that our approach is a reliable alternative for the calculation of photoelectron spectra in molecules. For the direct process, Fig.~\ref{Fig:SFAContributions}(b), we observe that both, the direct $\textit{Left}$ and direct $\textit{Right}$ terms, contribute within a comparable energy range. In addition, both terms show the same behavior, having exactly the same energy for the interference minimum: the coherent sum of these two terms, the total direct contribution (solid blue line), has a deeper minimum value around $1.0$~a.u. In Fig.~\ref{Fig:SFAContributions}(c) we observe that the local term (green doted line) contributes mostly in the low energy region of the ATI spectrum. Furthermore, the Non-Local and Cross terms do not contribute for electron energies $E_{p} \gtrsim6 U_p$. It is also demonstrated that the contribution of $|b_{Nl+Cross}(\textbf{p},t)|^2$ becomes even less important for larger internuclear distances as it is expected. Finally, in Fig.~\ref{Fig:SFAContributions}(d) we show that the Local Right and Local Left contributions have the same shape and contribute to the whole energy range.

\begin{figure}[htb]
            \subfigure[~Direct Term]{\includegraphics[width=0.45\textwidth] {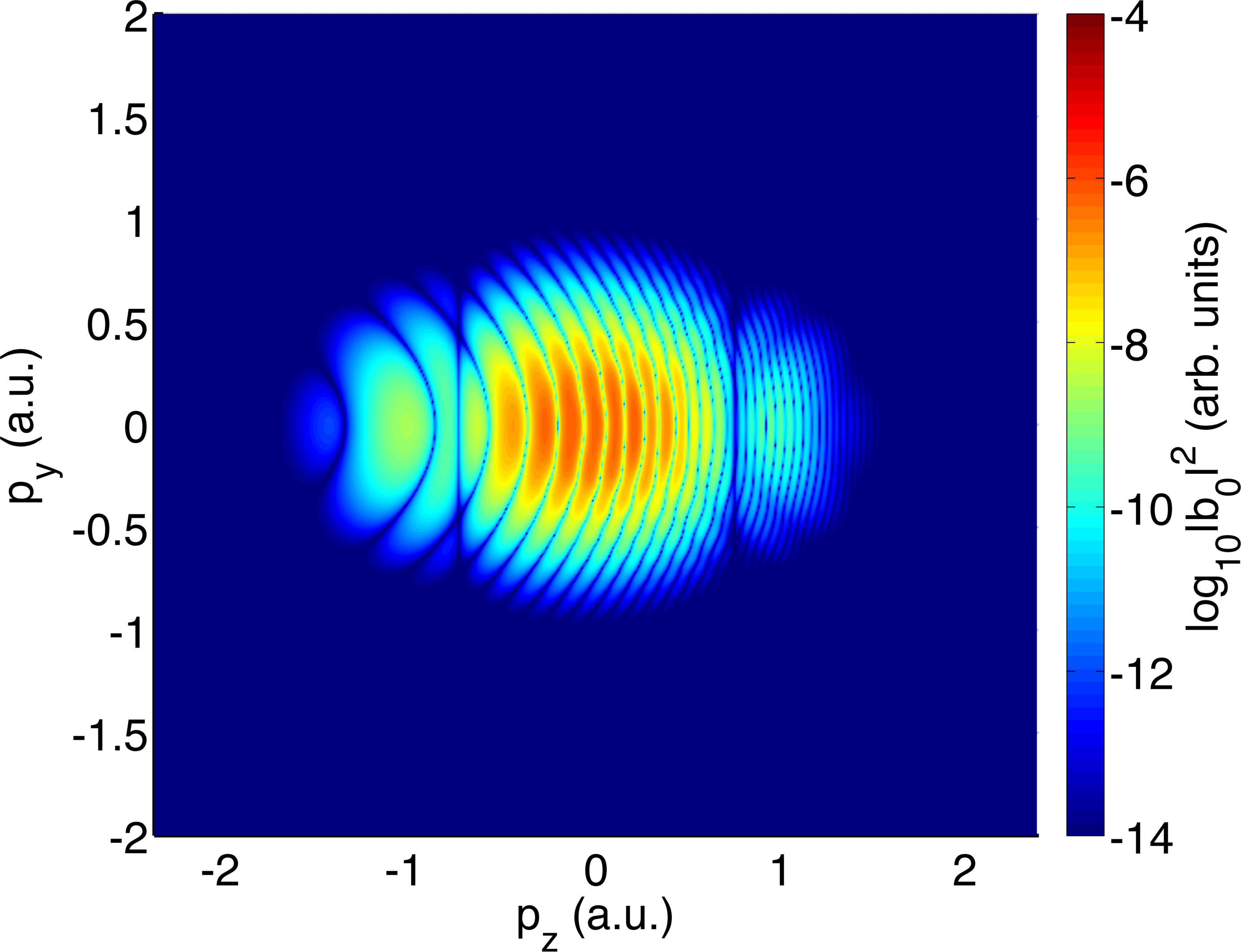}
            \label{fig:subfig}}
     \subfigure[~Non-Local and Cross Terms]{\includegraphics[width=0.45\textwidth]{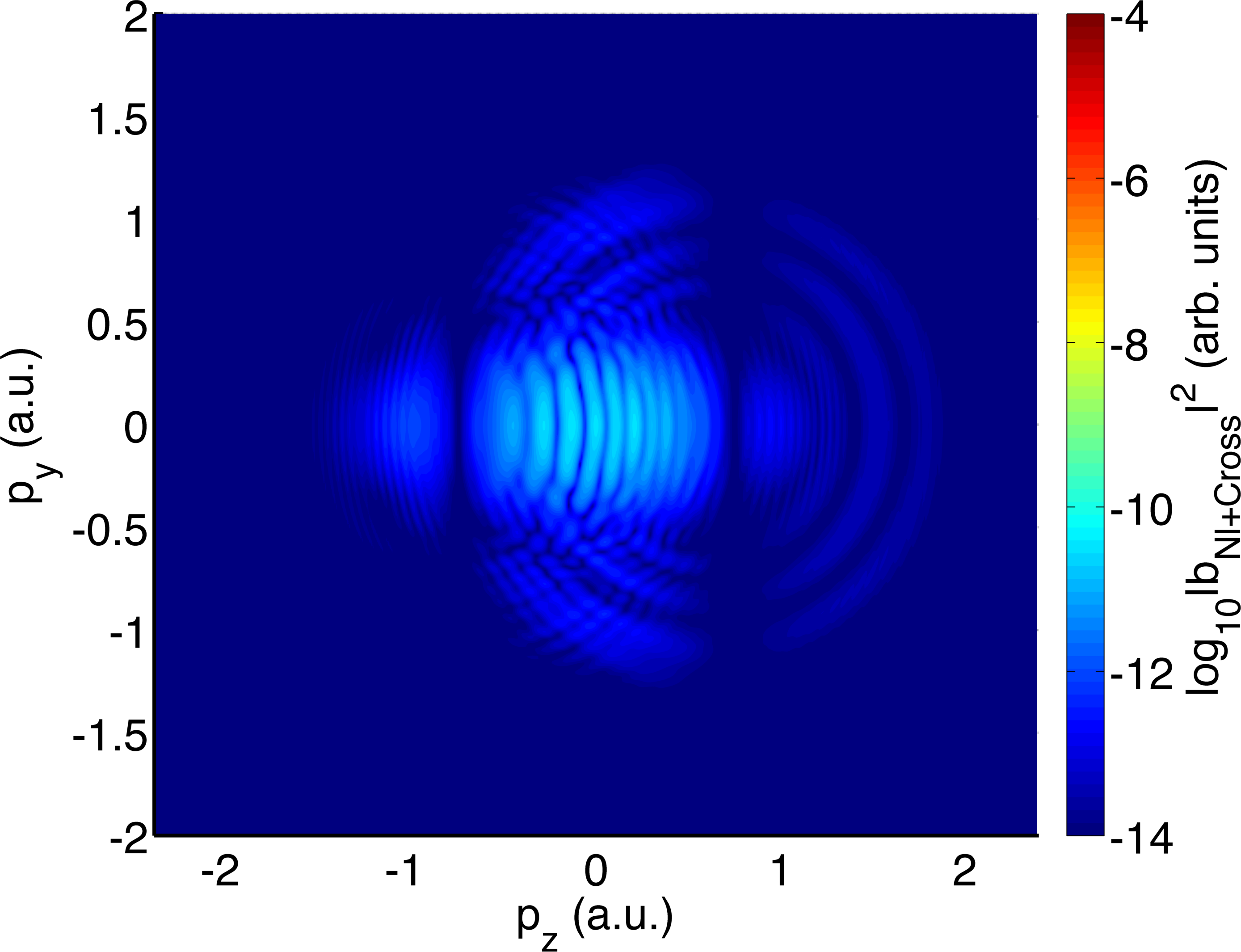}}
     \subfigure[~Local Term]{\includegraphics[width=0.45\textwidth]{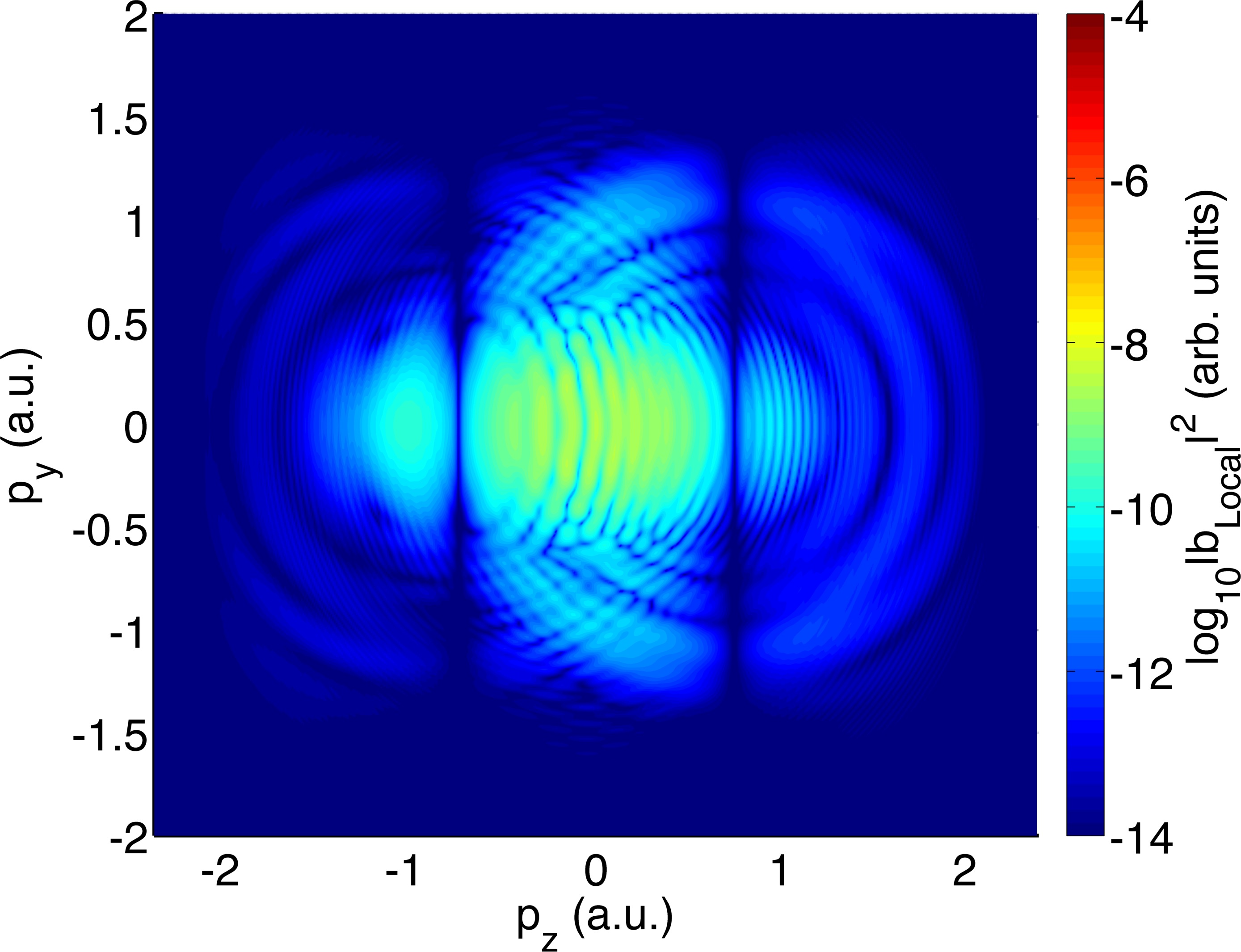}}
     \subfigure[~Total contribution]{\includegraphics[width=0.45\textwidth]{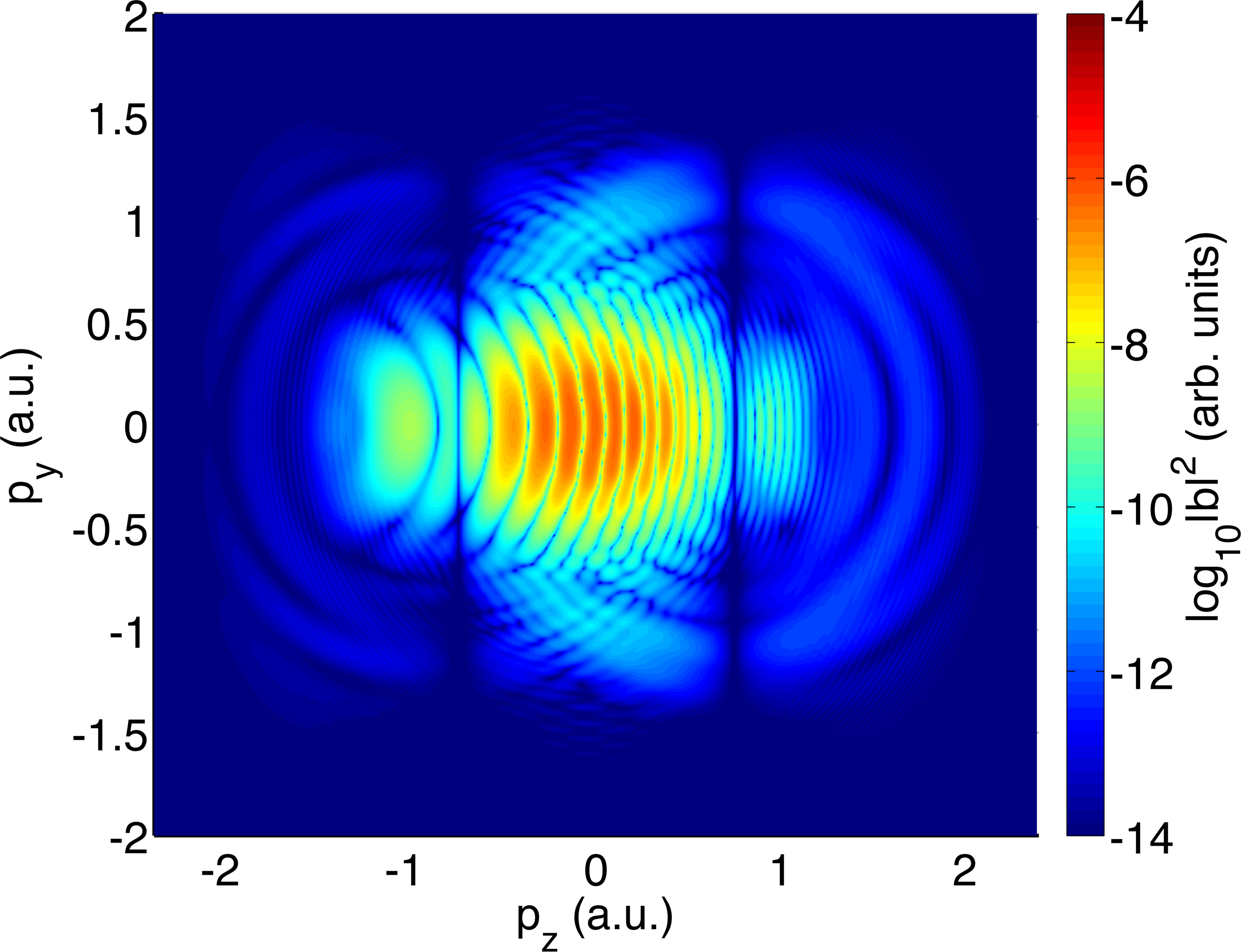}}
    \caption{(color online) Different contributions to the photoelectron spectra for a 2D-momentum plane $(p_z,p_y)$. ATI photoelectron spectra (in logarithmic scale) as a function of the momentum $(p_z,p_y)$ computed by our quasi-classical model for each term. (a) Direct term, (b) Non-Local and Cross terms, (c) Local term and (d) Total contribution. We use a laser field with a peak intensity of $I_0= 1\times10^{14}$~W$\,\cdot$\,cm$^{-2}$ and the internuclear distance is set to $R=4.2$~a.u.
 }
    \label{Fig:2DdTerm}
\end{figure}

Having in mind a deeper analysis of the ATI processes, we extend our numerical calculations from a 1D-momentum line (Fig.~\ref{Fig:SFAContributions}) to a 2D-momentum plane. The results of our computations are shown in Fig.~\ref{Fig:2DdTerm}. Here we depict the different contributions using our analytical quasi-classical  ATI model. For this calculation we use a laser field with a peak intensity of $I_0= 1\times10^{14}$~W$\,\cdot$\,cm$^{-2}$ and the internuclear distance is set to $R=4.2$~a.u.
The 2D calculations resembles the features of the 1D ones: the low momenta region of the spectrum is dominated by the direct process (Fig.~\ref{Fig:2DdTerm}(a)), meanwhile the rescattering term- dominated by the Local processes- is important for large electron momenta (Fig.~\ref{Fig:2DdTerm}(c)). Furthermore, the Non-Local and Cross (Fig.~\ref{Fig:2DdTerm}(b)) processes can be neglected when compared to the Local process. In all the figures we clearly distinguish the position of a deep minimum about $p=$ 0.74 a.u./$E=$1.23 $U_p$ and the well known asymmetric rings. Furthermore, we observe a symmetry of the structures about the $p_y$ axis for all the terms and a left-right asymmetric for electrons with $p_z<0$ or $p_z>0$. 
As was already mentioned, one of the advantages of our diatomic SFA model is that it allows us to account for the individual contributions to the ATI spectrum. In addition, besides of being analytically formulated, our model is able to switch on and off each of the ionization mechanisms which build up the final total and experimentally accessible ATI momentum spectrum $|b({p}_z,p_y,t_{\rm F})|^2$.

 As was mentioned at the outset, one of the main concerns with our model is to find a way to retrieve structural information of the molecular system starting from the ATI spectra. In the next we perform a detailed analysis of the interference pattern for different internuclear distances. Here, the well known two-slit interference formula, $ p= \frac{(2n+1)\pi}{R\:\cos\theta}$, \cite{CarlaPRA2007} is used in order to extract the internuclear distance from the interference pattern present in the photoelectron spectra. 
\begin{figure}[htb]
            \subfigure[$\,$ $R = 14.2$ a.u.]{ \includegraphics [width=0.45\textwidth] {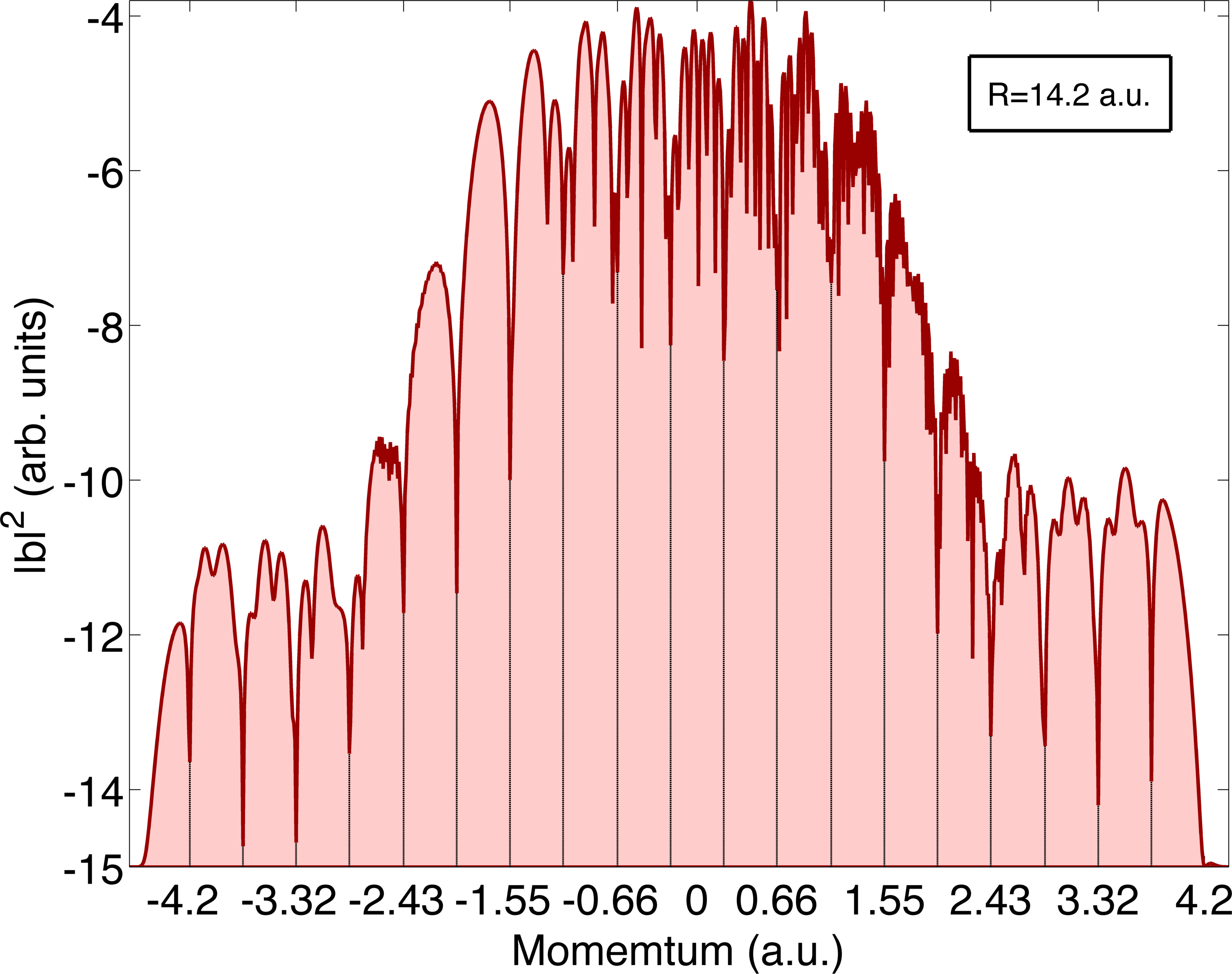}
            \label{fig:subfiga}}
                         \subfigure[$\,$Internuclear distance scan.]{\includegraphics[width=0.45\textwidth]{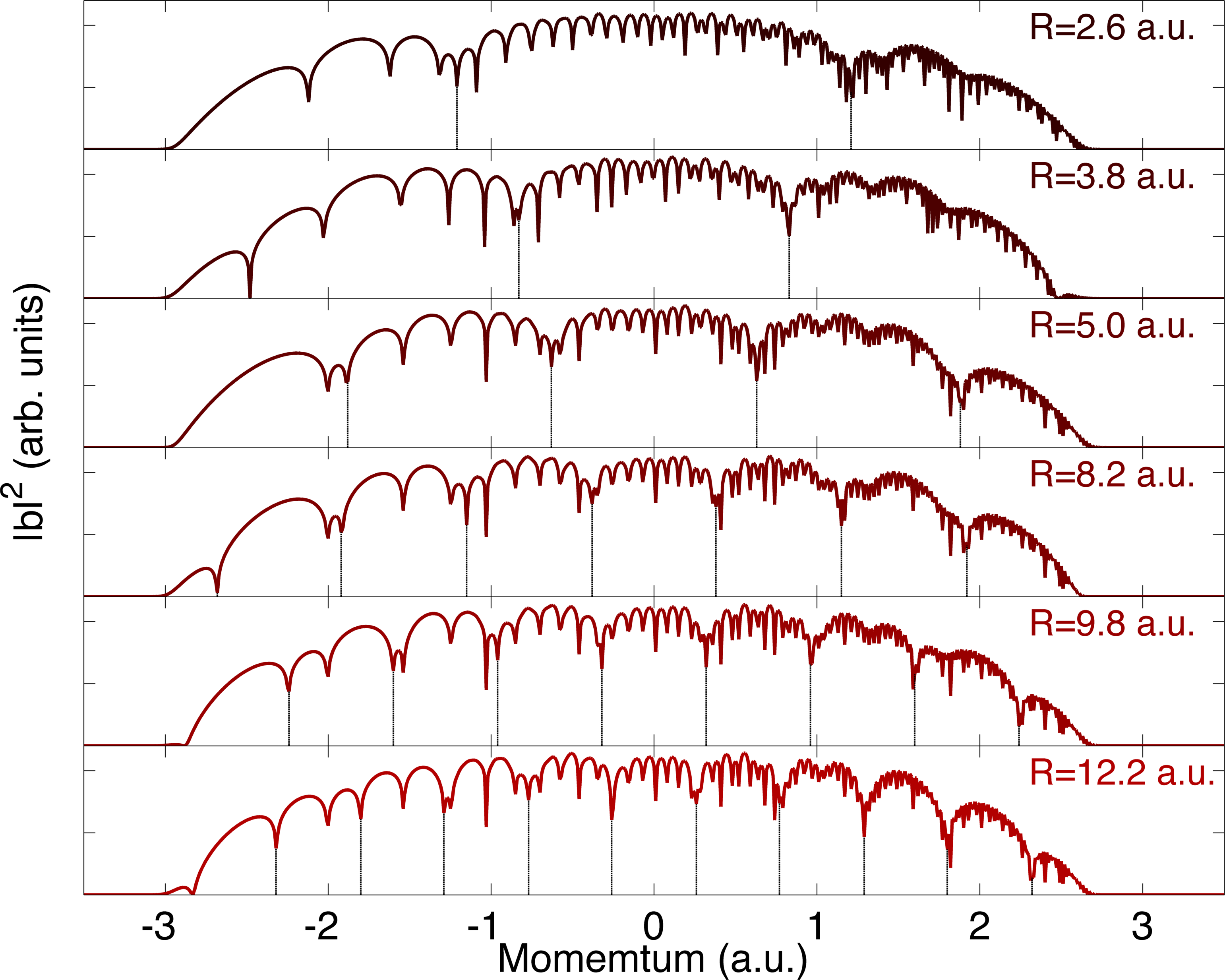}}
                  
                    \caption{(color online) ATI spectra calculated using the SFA model for an intensity value of $I_0= 4\times10^{14}$~W$\,\cdot$\,cm$^{-2}$, as a function of the momentum. (a) Photoelectron spectra computed for $R=14.2$ a.u.; (b) the same as (a) but for an internuclear range $R=[2.6,\,12.2]$~a.u. 
                                    }  
 \label{Fig:SFAInternDistance}
\end{figure}
In Fig.~\ref{Fig:SFAInternDistance} we show the photoelectron distribution or ATI spectra ($|b(p,t)|^2$), in logarithmic scale, as a function of the final electron momentum for different internuclear distances. Figure~\ref{Fig:SFAInternDistance}(a) depicts the ATI spectra for a large value of the internuclear distance: $R = 14.2$ a.u.,~in order to see a considerable number of interference minima. Furthermore,  the ATI spectra in Fig.~\ref{Fig:SFAInternDistance}(b) is computed varying the internuclear distance $R$ (see the panel labels for the values). The dashed black lines represent the expected minima calculated by applying the two-slit interference formula. As it is observed, our model is able to reproduce all the interference minima and this is a clear evidence that the photoelectron spectra contains structural information of the molecular system. 

\subsection{Theory vs experimental results}

In order to conclude our analysis and as an additional validation of our model, we compare the results computed using the SFA approach with experimental data obtained at ICFO for O$_2^+$ molecules~\cite{Pullen2016}.  The experimental data were taken for randomly oriented molecules and the laser pulse was CEP randomized, i.e.~an average of the theoretical results over both the molecular orientations and different CEP values is required for an accurate theoretical description. The reported laser peak intensity and wavelength are $I_0= 8.5\times10^{13}$~W$\,\cdot$\,cm$^{-2}$ and $\lambda=3.1$ $\mu$m, respectively. The laser pulse has a duration of 75 fs full width at half maximum at a repetition rate of 160 kHz. Furthermore, the O$-$O bond length is retrieved from the photoelectron spectra and set  to a value $R=1.17$ \AA\, (2.21 a.u.)~\cite{Pullen2016}. 
This value of $R$ corresponds to an ionization potential energy of $I_p=0.93$ a.u and, in order to reproduce this value, in our model we set the parameters of the non-local potential $\Gamma=0.75$ and $\gamma=0.097$~a.u. With these values we obtain a very good fit to the dissociation energy, $E_d=18.5$ eV, and the equilibrium internuclear distance, $R=1.116$ {\rm \AA} (2.11 a.u.), reported in the literature~\cite{O2plus}. 

\begin{figure}[htb]
            \subfigure[$\,$SFA]{ \includegraphics [width=0.85\textwidth] {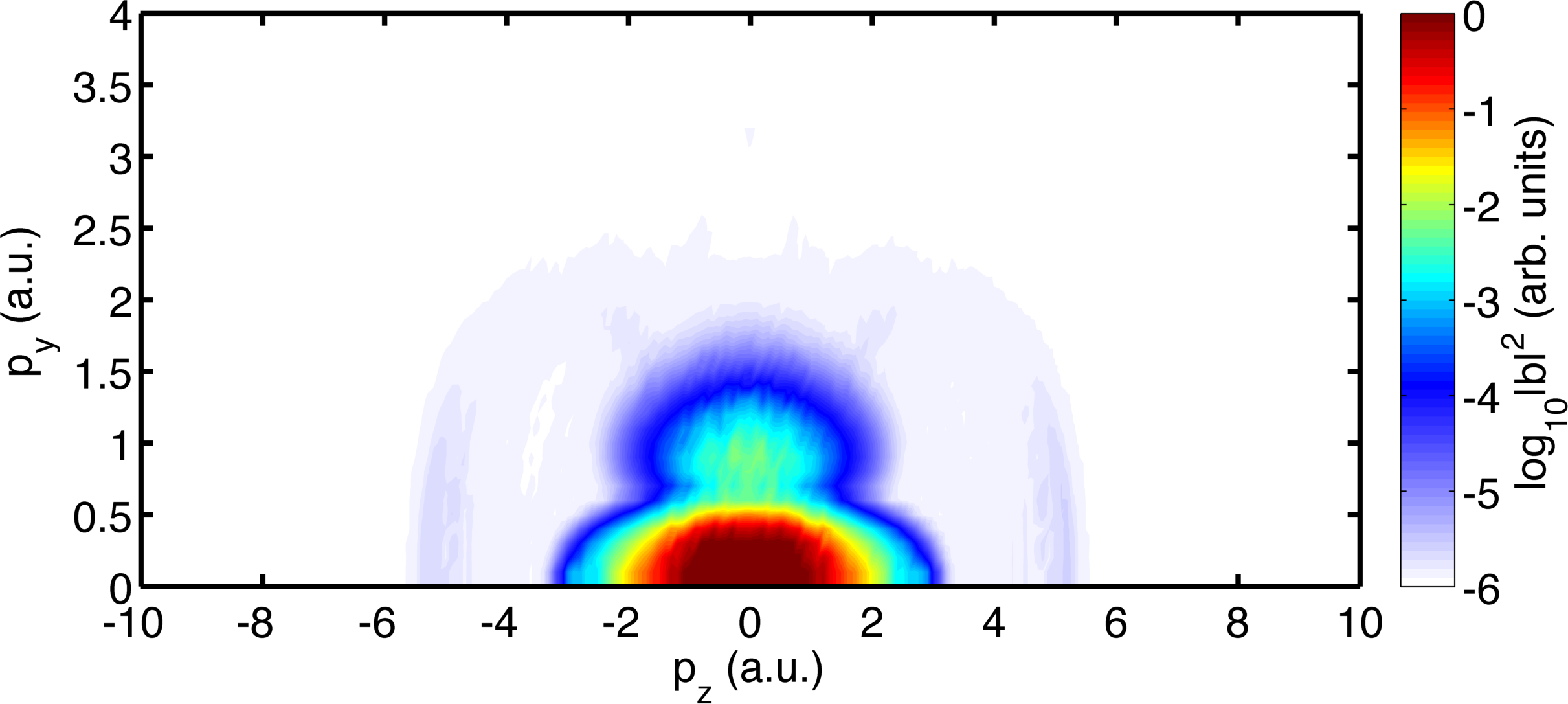}
            \label{fig:subfiga}}
                         \subfigure[$\,$Experiment]{\includegraphics[width=0.85\textwidth]{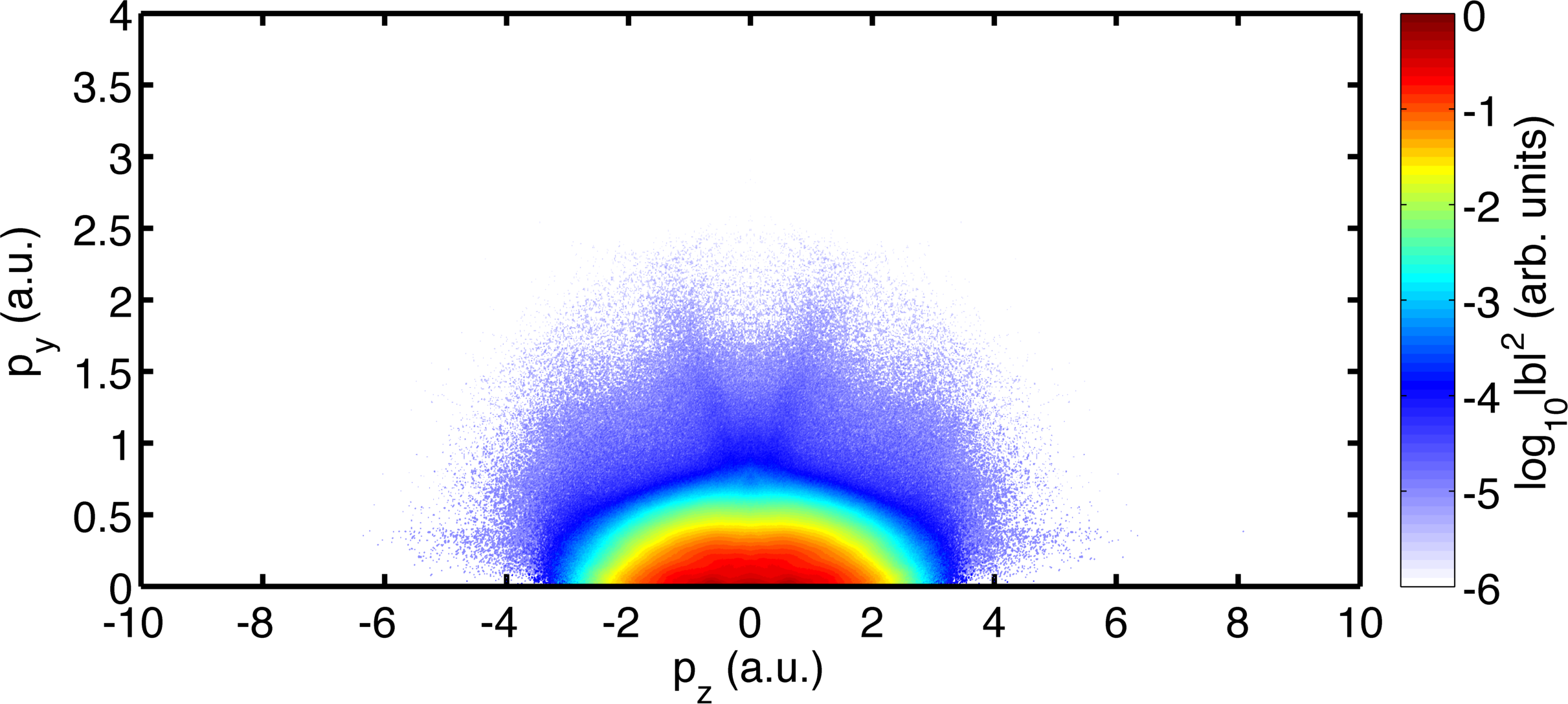}}
                  
                    \caption{(color online) Photoelectron spectra for the O$_2^+$ molecule. (a) ATI spectra calculated using the SFA model; (b) Experimental ATI spectra obtained in the Attosecond and Ultrafast Optics group at ICFO~\cite{Pullen2016}. In the theoretical calculations the laser peak intensity, wavelength and total time duration are $I_0= 8.5\times10^{13}$~W$\,\cdot$\,cm$^{-2}$, $\lambda=3.1 \mu$m and 52 fs (5 optical cycles), respectively. In addition, the internuclear distance is set to $R=2.21$~a.u. (1.17 \AA) (see the text for details).}                                      
 \label{Fig:SFAVsExperiment}
\end{figure}

In order to accurately compare with the experimental measurements, the calculated ATI spectra are averaged over the orientation of the molecule with respect to the laser-polarization axis, using 8 values from $\theta = [0^{\circ}-360^{\circ}]$. In addition, an average over the CEP values, and for the same orientation range, is considered. For symmetry considerations, only 16 different sets of photoelectron spectra are computed. For each set a total of 8192 points in the $(p_z,p_y)$ plane are used. Around 150,000 CPU-hours were employed for the whole ATI computation. 
For the comparison experiment vs. theory, we employ the same laser peak intensity (no focal averaging is considered in the calculations) and the internuclear distance reported in the experiment for each calculation. The result of this comparison is depicted in Fig.~\ref{Fig:SFAVsExperiment}.

In Fig.~\ref{Fig:SFAVsExperiment} (a) we show the calculated total ionization probability $|b(\textbf{p},t)|^2$, Eq.~(\ref{Eq:b_T}) and in Fig.~\ref{Fig:SFAVsExperiment} (b) we present the experimental data. In order to make an easier comparison the theoretical calculations are multiplied by a constant factor.
The plots show that our model is in very good agreement with the experimental measurements. In fact, both panels present the same  colour scales, covering six orders of magnitude.  Both the simulated and measured data exhibit the same regions of signal, with comparable amplitudes for all the longitudinal momentum. In addition a similar dome structure, around $p_z=[-3; 3]$ and $p_y=[0; 0.7]$, is observed in both pictures. Note that the interference fringes are not observed neither in the experimental nor in the theoretical calculations. Thereby the different recollision scenarios in terms of electron trajectories are washed out due to both the molecular orientation and CEP averages.


\section{Conclusions and Outlook}

We have presented a novel, simple and analytical model to describe the above-threshold ionization (ATI) process from a diatomic molecule while an ultra-intense infrared laser field drives the system. Our approach is based on the analytical solution of the time dependent Schr\"odinger equation by means of considering the bound and scattering states as a composition of two states depending on the relative position of the atoms inside of the molecule, within the framework of both the SFA and SAE approximations. Further, a systematic and analytical way for computing both the bound-free dipole with respect to each center and the rescattering transition matrix elements is developed. This is the advantage of our theoretical model with respect to those used before, since it gives a solution free of any artificial and nonphysical effects. In fact a correct asymptotic behaviour for ${\textbf R}\to \infty$ and yet to ${\textbf R}\to 0$ is obtained. In addition, the rescattering transition amplitude is written as a sum of components, obtained from equations which describe each rescattering process (Local, Non-Local and Cross) independently. 

Our model is an extension and generalization of previous works for atomic systems. It shows that each component contributes, in a different way, to a different region of the ATI spectrum. The results indicate, as expected, that the main contribution to the rescattering transition amplitude corresponds to Local events, with the Cross and Non-Local terms playing almost no role, when the internuclear distance becomes much larger than the equilibrium one. We should stress, however, that our model is by no means a simple correction to the well established SFA. Here we do predict physical processes for the first time, i.e.~those modelled by the Non-Local and Cross terms, and we do provide methods to identify their contributions, which, at the end can be quite significant for small internuclear distances compared to the equilibrium one. On top of that, our version of SFA compares very well with the TDSE and the experimental data, which provides an even stringent argument in favour of our model. 

In this paper we used our model for a proof-of-principle that photoelectron spectra contains structural information about the target system. We employ it to retrieve, satisfactorily, the internuclear distance of a H$_2^+$ molecule using a simple interference equation. The magnitude of the ionization probability for different values of $R$ was calculated and compared with the TDSE outcomes, and the results present a very good agreement.  Both models exhibit the similar behavior: the ionization probability shows the same tendencies, namely it starts to increase linearly with $R$ reaching a saturation value from which it remains constant. The comparison TDSE with SFA, as well as  the good agreement between the experiment and the simulations, validates our theoretical model and allow us to believe that extensions to more complicated systems, such as polyatomic molecules with more than three centres, are perfectly feasible. We hope that our work paves the way toward fascinating studies of structural information and charge migrations in the fragmentation processes in large molecules and other complex targets.

\acknowledgments{This work was supported by the project ELI--Extreme Light Infrastructure--phase 2 (CZ.02.1.01/0.0/0.0/15\_008/0000162 ) from European Regional Development Fund, Ministerio de
Econom\'{\i}a y Competitividad through Plan Nacional
(FIS2011-30465-C02-01, FrOntiers of QUantum Sciences (FOQUS):
Atoms, Molecules, Photons and Quantum Information
FIS2013-46768-P, FIS2014-56774-R, and Severo Ochoa
Excellence Grant SEV-2015-0522),  the Catalan Agencia de Gestio d'Ajuts
Universitaris i de Recerca (AGAUR) with SGR 2014-2016, Fundaci\'o
Privada Cellex Barcelona and funding from the European Union’s Horizon 2020 research and innovation programme under the Marie Sklodowska-Curie grant agreement No. 641272 and Laserlab-Europe (EU-H2020 654148). N.S. was supported by the Erasmus Mundus Doctorate Program
Europhotonics (Grant No. 159224-1-2009-1-FR-ERA MUNDUS-EMJD). N. S., A.
C. and M. L. acknowledge ERC AdG OSYRIS and EU FETPRO QUIC. J.B. acknowledges
FIS2014-51478-ERC.}

\appendix
\section{Treatment of the asymptotic behaviour when: ${\textbf R}\to 0$ }

In this Appendix we prove, analytically, that our model is capable to satisfy the asymptotic limit when the separation between the two atoms of the molecule is close to zero. For this condition our model describes a single atom that satisfies the equations previously presented in Ref.~\cite{PRANoslen2015}. In this sense our theoretical formulation for a diatomic molecule remains compatible with the atomic model.

\subsection{Bound States and Bound-Continuum transition matrix element}

When the internuclear distance is close to zero the bound state of our diatomic molecule is equal to the bound state of an atom, $\lim_{\textbf{R} \to 0} \Psi_{0\textbf{M}}(\textbf{p})= \Psi_0(\textbf{p})$. The wave function describing the bound state for the atomic system~\cite{PRANoslen2015} reads as:
  \begin{eqnarray}
 \Psi_0(\textbf{p}) &=& \frac{\mathcal{N}}{\sqrt{(p^2 + \Gamma ^2)}(\frac{p^2}{2} +I_p)},
 \label{Eq:WF1}
\end{eqnarray}
where, $\mathcal{N} = \bigg(\frac{\sqrt{2I_p}\big( \Gamma +  \sqrt{2I_p} \big)^2}{4\pi^2} \bigg)^{1/2}$, is the normalization constant. In order to perform the limit $\textbf{R} \to 0$ for the molecular bound-state we are going to use Eq.~(\ref{Eq:BSMolec}). If we write it as an explicit function of $\textbf{R}$ and taking the limit we have: 
 \begin{eqnarray}
\lim_{\textbf{R} \to 0} \Psi_{0\textbf{M}}(\textbf{p}) &=&\lim_{\textbf{R} \to 0}\Bigg\{\frac{ \mathcal{M} \: e^{\frac{ i \textbf{R}}{2} \cdot \textbf{p} } }{ \sqrt{(p^2 + \Gamma^2})(\frac{p^2}{2} + I_p)}   +   \frac{ \mathcal{M} \: e^{-\frac{ i \textbf{R}}{2} \cdot \textbf{p} } }{ \sqrt{(p^2 + \Gamma^2})(\frac{p^2}{2} + I_p)} \Bigg\},\nonumber\\
&=&\frac{ \lim_{\textbf{R} \to 0} 2 \mathcal{M} }{ \sqrt{(p^2 + \Gamma^2})(\frac{p^2}{2} + I_p)}.
\label{Eq:BfLimit}
\end{eqnarray}

On the other hand, the normalization constant for the molecular bound state is given by Eq. (\ref{Eq:BNormConst}) and its limit is: 
\begin{eqnarray}
\lim_{\textbf{R} \to 0} 2\mathcal{M}  & = & \lim_{\textbf{R} \to 0} \: \frac{ 1} {\Bigg(  \frac{ 2\pi^2}{( 2I_p-\Gamma^2 )^2} \Bigg\{  \frac{2\: e^{-R \Gamma}} {R}  -  \frac{2 \: e^{-R \sqrt{2I_p}}} {R} \Bigg( \frac{2 \sqrt{2I_p} + R(2I_p -\Gamma^2)}{2\sqrt{2I_p}} \Bigg) + \frac{(\sqrt{2I_p} - \Gamma)^2}{\sqrt{2I_p}}   \Bigg\} \Bigg)^{1/2} },\nonumber\\
 &= & \lim_{\textbf{R} \to 0} \:  \frac{ 1} {\Bigg( \frac{ 2\pi^2}{( 2I_p-\Gamma^2 )^2} \Bigg\{  2\: (-\Gamma + \sqrt{2I_p})  - \frac{2I_p - \Gamma^2}{\sqrt{2I_p}}+ \frac{(\sqrt{2I_p} -\Gamma)^2}{\sqrt{2I_p}}   \Bigg\} \Bigg)^{1/2} },\nonumber\\
& = &  \Big( \frac{\sqrt{2I_p}( \sqrt{2I_p} + \Gamma )^2} { 4 \pi^2 }\Big)^{1/2}=\mathcal{N},
\label{Eq.limitNCont}
\end{eqnarray}
\newline
from where the relation $\lim_{\textbf{R} \to 0} 2 \mathcal{M} =  \mathcal{N}$ is demonstrated. 

For the bound-continuum transition matrix element we follow the same analysis. By taking the asymptotic limit as:
\begin{eqnarray} 
\lim_{\textbf{R} \to 0}\textbf{d}_{m}( \textbf{p}_0) &=&\lim_{\textbf{R} \to 0}\Big\{ - 2\textit{i}\: \mathcal{M} \mathcal{A}( \textbf{p}_0) \big[ e^{\frac{ i \textbf{R}}{2} \cdot \textbf{p}_0} + e^{-\frac{ i \textbf{R}}{2} \cdot \textbf{p}_0 }  \big] \Big\},\nonumber\\
&= &\textit{i} \: \textbf{p}_0 \frac{(p_0^2 +\Gamma^2)+( \frac{p_0^2}{2} + I_p )}{(p_0^2 + \Gamma^2)^{\frac{3}{2}}(\frac{p_0^2}{2} + I_p)^2} \: \lim_{\textbf{R} \to 0} 2 \mathcal{M},
\end{eqnarray}
and using Eq. (\ref{Eq.limitNCont}), we obtain:
 \begin{equation}
\lim_{\textbf{R} \to 0}\textbf{d}_{m}( \textbf{p}_0) = \textit{i} \: \textbf{p}_0 \frac{(p_0^2 +\Gamma^2)+( \frac{p_0^2}{2} + I_p )}{(p_0^2 + \Gamma^2)^{\frac{3}{2}}(\frac{p_0^2}{2} + I_p)^2} \: \mathcal{N},
\label{Eq:dpLimt}
\end{equation}
which is exactly the bound-continuum transition matrix element for the atomic system, i.e.~:
 \begin{equation}
\textbf{d}( \textbf{p}_0) = \textit{i}\mathcal{N} \textbf{p}_0 \frac{(p_0^2 + \Gamma^2) + (\frac{p_0^2}{2} + I_p)}{(p_0^2 + \Gamma^2)^{\frac{3}{2}}(\frac{p_0^2}{2} + I_p)^2}.
\label{Eq.Atomicdp}
\end{equation}

\subsection{Scattering states and Continuum-Continuum transition matrix element}

For the scattering states we will prove that: $\lim_{\textbf{R} \to 0} \Psi_{\textbf{M} \textbf{p}_0}(\textbf{p}) = \Psi_{\textbf{p}_0}(\textbf{p})$. For the atomic system  the scattering state obeys the equation:
\begin{eqnarray}
\Psi_{\textbf{p}_0}(\textbf{p})  = \delta(\textbf{p}-\textbf{p}_0) +  \frac{\mathcal{B}({\bf p}_0)}{\sqrt{p^2 + \Gamma ^2}\bigg(p_0^2 - p^2 +\textit{i}\epsilon \bigg)},
\label{Eq:RescA}
\end{eqnarray}
 where, $\mathcal{B}(\textbf{p}_0)  = -\frac{2\gamma}{(p_0^2 +\Gamma^2)^{\frac{1}{2}}}\bigg(1- \frac{4\pi^2\textit{i}\gamma}{| \textit{p}_0| +\textit{i}\Gamma }\bigg)^{-1}$, is the normalization constant.
 \newline
 
From Eqs.(\ref{Eq.ScatState}) and (\ref{Eq:RescM}) the asymptotic limit for the molecular system reads as:  
\begin{eqnarray}
\lim_{\textbf{R} \to 0} \Psi_{\textbf{M} \textbf{p}_0}(\textbf{p})&= &  \delta(\textbf{p}-\textbf{p}_0) + \lim_{\textbf{R} \to 0} \Bigg\{ \frac{\mathcal{D}_1(\textbf{p}_0) \big[ e^{-\frac{ i \textbf{R}}{2} \cdot (\textbf{p} - \textbf{p}_0) }+\: e^{\frac{ i \textbf{R}}{2} \cdot (\textbf{p} - \textbf{p}_0)} \big]}{ \sqrt{p^2 + \Gamma^2} \:(p_0^2 -p^2 +\textit{i}\epsilon) }\nonumber\\
&&- \frac{\mathcal{D}_2 (\textbf{p}_0) \big[ e^{-\frac{ i \textbf{R}}{2} \cdot (\textbf{p} + \textbf{p}_0) } +e^{\frac{ i \textbf{R}}{2} \cdot (\textbf{p} + \textbf{p}_0) }\big]} { \sqrt{p^2 + \Gamma^2} \:(p_0^2 -p^2  +\textit{i}\epsilon) } \Bigg\},\nonumber\\
&= &\delta(\textbf{p}-\textbf{p}_0) + \frac{2 \: \lim_{\textbf{R} \to 0} \Big\{ \mathcal{D}_1(\textbf{p}_0)  - \mathcal{D}_2(\textbf{p}_0) \Big\} }{ \sqrt{p^2 + \Gamma^2} \:(p_0^2 -p^2 +\textit{i}\epsilon) }  .
\end{eqnarray}
\newline
Working with the above equation we are going to prove that:
  \begin{equation}
\mathcal{B}(\textbf{p}_0)=2\:\lim_{\textbf{R} \to 0}  \Big\{ \mathcal{D}_1(\textbf{p}_0) - \mathcal{D}_2(\textbf{p}_0)  \Big \}.
\end{equation}  
Substituting the values of the constants we have,   
\begin{eqnarray}
\mathcal{B}(\textbf{p}_0)&=&\frac{2\gamma}{\sqrt{p_0^2 + \Gamma^2}}\lim_{\textbf{R} \to 0}  \Big\{ \frac{ 1 +I_1 }{I^2_2 - [1 + I_1 ]^2}  - \frac{I_2}{I^2_2 - [1 + I_1 ]^2} \Big \}.
\end{eqnarray}
By taking the limit $\textbf{R} \to 0$ of $I_2$ and using Eq.~(\ref{Eq:I2}):
\begin{eqnarray}
\lim_{\textbf{R} \to 0} = \frac{- 2 \pi^2 \:\gamma} {R\:(p_0^2 + \Gamma^2  + i\epsilon)} \bigg[ e^{iR\:\sqrt{p_0^2 +i\epsilon}} - e^{-R\: \Gamma} \bigg ]= \frac{- 2\pi^2\:\gamma  }{\Gamma - i\sqrt{|p_0^2 + \textit{i}\:\epsilon|}},
\end{eqnarray}
we find that $\lim_{\textbf{R} \to 0} I_2=I_1$. From this last result we can write,
\begin{eqnarray}
\mathcal{B}(\textbf{p}_0)&=&\frac{2\gamma}{\sqrt{p_0^2 + \Gamma^2}}  \Big[ \frac{ 1}{I^2_1 - [1 + I_1]^2} \Big ]=\frac{-2\gamma}{\sqrt{p_0^2 + \Gamma^2}}  [ 1+ 2I_1]^{-1}, \nonumber\\
&=&\frac{-2\gamma}{\sqrt{p_0^2 + \Gamma^2}}  \Bigg( 1- \frac{4\pi^2\textit{i}\gamma}{| \textit{p}_0| +\textit{i}\Gamma} \Bigg)^{-1},
\label{Eq:ScNorConst}
\end{eqnarray}
which is identical to Eq.~(33) of Ref.~\cite{PRANoslen2015}.
\newline
Concluding the analysis of the diatomic molecular system when $\textbf{R} \to 0$ we proceed to demonstrate that the continuum-continuum molecular matrix element is equal to the continuum-continuum atomic matrix element. The dipole matrix element for the atomic system can be written as:
\begin{equation}
 \textbf{g}(\textbf{p}_1, \textbf{p}_2) =\textit{i}\mathcal{B}(\textbf{p}_2)\textbf{p}_1\Bigg\{\frac{3p_1^2 - p_2^2 +2\Gamma^2 }{(p_1^2 +\Gamma ^2)^{\frac{3}{2}}(p_1^2 - p_2^2 + \textit{i}\epsilon)^2}\Bigg\} -\textit{i}\mathcal{B}^*(\textbf{p}_1)\textbf{p}_2\Bigg\{\frac{3p_2^2 - p_1^2 +2\Gamma^2 }{(p_2^2 +\Gamma ^2)^{\frac{3}{2}}(p_1^2 - p_2^2 +\textit{i}\epsilon)^2}\Bigg\}.
 \end{equation}

Taking the limit in Eq. (\ref{Eq:gp1p2M}):
\begin{eqnarray}
\lim_{\textbf{R} \to 0} \textbf{g}_m( \textbf{p}_1,\textbf{p}_2)& = &2 \lim_{\textbf{R} \to 0} \Bigg[\mathcal{Q}_1(\textbf{p}_1, \textbf{p}_2)\:  +  \mathcal{Q}_2 (\textbf{p}_1, \textbf{p}_2) \: \Bigg],\nonumber\\
& = &i \mathcal{C}_1 (\textbf{p}_1, \textbf{p}_2 ) \:2 \lim_{\textbf{R} \to 0} \Bigg[\frac{ 1 +I_1 }{I^2_2 - [1 + I_1 ]^2}  - \frac{I_2}{I^2_2 - [1 + I_1 ]^2} \Bigg]_{p_2} \nonumber\\
&&- i \mathcal{C}_2 (\textbf{p}_1,\textbf{p}_2) \:2\lim_{\textbf{R} \to 0} \Bigg[\frac{ 1 +I_1 }{I^2_2 - [1 + I_1 ]^2}  - \frac{I_2}{I^2_2 - [1 + I_1 ]^2} \Bigg]^*_{p_1},
\end{eqnarray} 
where,
\begin{equation}
\begin{split}
\lim_{\textbf{R} \to 0} \textbf{g}_m( \textbf{p}_1,\textbf{p}_2) = & i\: \mathcal{C}_1 (\textbf{p}_1,\textbf{p}_2)\Bigg\{ \frac{2\gamma}{\sqrt{p_2^2 + \Gamma^2}}\lim_{\textbf{R} \to 0}  \Bigg [ \frac{ 1}{I^2_1 - [1 + I_1 ]^2} \Bigg ]_{p_2} \Bigg\} \\
&-i\: \mathcal{C}_2 (\textbf{p}_1, \textbf{p}_2 )\Bigg\{ \frac{2\gamma}{\sqrt{p_1^2 + \Gamma^2}}\lim_{\textbf{R} \to 0}  \Bigg [ \frac{ 1}{I^2_1 - [1 + I_1 ]^2} \Bigg ]^*_{p_1} \Bigg\}.
\end{split}
\end{equation}
  
By following the same procedure as in Eq.~(\ref{Eq:ScNorConst}), we finally obtain that:

 \begin{equation}
\mathcal{B}(\textbf{p}_2)=\Bigg\{ \frac{2\gamma}{\sqrt{p_2^2 + \Gamma^2}}\lim_{\textbf{R} \to 0}  \Bigg [ \frac{ 1}{I^2_1 - [1 + I_1 ]^2} \Bigg ]_{p_2} \Bigg\} ,
\end{equation}
and 
 \begin{equation}
\mathcal{B}^*(\textbf{p}_1) =\Bigg\{ \frac{2\gamma}{\sqrt{p_1^2 + \Gamma^2}}\lim_{\textbf{R} \to 0}  \Bigg [ \frac{ 1}{I^2_1 - [1 + I_1 ]^2} \Bigg ]^*_{p_1} \Bigg\}.
\end{equation}  
 
Grouping conveniently the above equation we get:
\begin{equation}
\lim_{\textbf{R} \to 0} \textbf{g}_m( \textbf{p}_1,\textbf{p}_2) =\textit{i}\mathcal{B}(\textbf{p}_2)\textbf{p}_1\Bigg\{\frac{3p_1^2 - p_2^2 +2\Gamma^2 -i\epsilon}{(p_1^2 +\Gamma ^2)^{\frac{3}{2}}(p_1^2 - p_2^2 + \textit{i}\epsilon)^2}\Bigg\} -\textit{i}\mathcal{B}^*(\textbf{p}_1)\textbf{p}_2\Bigg\{\frac{3p_2^2 - p_1^2 +2\Gamma^2 +i\epsilon}{(p_2^2 +\Gamma ^2)^{\frac{3}{2}}(p_1^2 - p_2^2 +\textit{i}\epsilon)^2}\Bigg\},
 \end{equation}  
which is nothing else that the atomic transition matrix continuum-continuum element (see Eq.~(37) of Ref.~\cite{PRANoslen2015}).
 
We have indeed demonstrated, with the above analysis and relations, that the theoretical model presented in this contribution configures a general model which not only describes the ATI process in diatomic molecules, but is also able, when the appropriate limits are taken, to model the atomic ATI. 

%


\end{document}